\documentclass[journal]{IEEEtran}
\usepackage{cite}
\usepackage{graphicx}
\usepackage{epstopdf}
\usepackage[subrefformat=parens, labelformat=parens]{subfig}
\usepackage[font=footnotesize, labelfont=footnotesize]{caption}
\usepackage{caption}
\usepackage{times}
\usepackage{siunitx} 
\usepackage{color}
\usepackage{lipsum}
\usepackage{algorithm, algorithmicx, algpseudocode}
\usepackage{amsmath, amssymb}
\usepackage{upgreek}
\usepackage{bm, bbold}
\usepackage{tabularx}
\usepackage{enumitem}
\usepackage{flushend}

\newtheorem{theorem}{Theorem}
\newtheorem{definition}{Definition}

\newtheorem{proposition}{Proposition}
\newtheorem{lemma}{Lemma}

\DeclareSIUnit \uAhr{ \micro Ahr}
\DeclareSIUnit \Kbps{ Kbps}
\DeclareSIUnit \Mbps{ Mbps}
\DeclareSIUnit \dBm{ dBm}

\usepackage[normalem]{ulem}

\begin{document}

\setlength{\abovedisplayskip}{3pt}
\setlength{\belowdisplayskip}{3pt}

\sloppy

\title{Maximizing Broadcast Throughput \\Under Ultra-Low-Power Constraints}

\author{Tingjun~Chen, 
        Javad~Ghaderi, 
        Dan~Rubenstein, 
        and~Gil~Zussman
\thanks{This research was supported in part by ARO grant 9W911NF-16-1-0259, NSF grant ECCS-1547406, and the People Programme (Marie Curie Actions) of the European Union's Seventh Framework Programme (FP7/2007-2013) under REA grant agreement n${^{\text{o}}} $[PIIF-GA-2013-629740].11. A partial and preliminary version of this paper appeared in ACM CoNEXT'16, Dec. 2016~\cite{chen_conext2016}, and some results were presented in IEEE WCNC'17 Workshop on Energy Harvesting and Remotely Powered Wireless Communications for the IoT (invited), Mar. 2017.}
\thanks{T. Chen, J. Ghaderi, and G. Zussman are with the Department of Electrical Engineering, Columbia University, New York, NY, USA (email: \{tingjun, jghaderi, gil\}@ee.columbia.edu).}
\thanks{D. Rubenstein is with the Department of Computer Science, Columbia
University, New York, NY 10027, USA (e-mail: danr@cs.columbia.edu).}
}

\maketitle


\newcommand{\name}{\textsf{\small EconCast}}
\newcommand{\namecaption}{\textsf{\footnotesize EconCast}} 
\newcommand{\namecap}{\textsf{\small \name-C}}
\newcommand{\namecapcaption}{\textsf{\footnotesize EconCast-C}} 
\newcommand{\namenoncap}{\textsf{\small \name-NC}}

\newcommand{\namelp}{(\textsf{P1})} 
\newcommand{\namecp}{(\textsf{P4})} 
\newcommand{\namelpgput}{(\textsf{P2})} 
\newcommand{\namelpaput}{(\textsf{P3})} 

\newcommand{\removeeqnheader}{\vspace{-.1in}}

\newcommand{\setnodes}{{\cal N}}  
\newcommand{\setnode}{{\cal N}}  
\newcommand{\numnode}{N}  
\newcommand{\setnodewoi}{\setnodes_{\overline{i}}}  
\newcommand{\setnodewoj}{{\setnodes_{\overline{j}}}}  
\newcommand{\neighborofi}{\setnodes(i)}  

\newcommand{\energybound}[1]{\rho_{#1}}  
\newcommand{\energyboundvec}{\bm{\uprho}}

\newcommand{\listencost}[1]{L_{#1}}
\newcommand{\listencostvec}{\mathbf{L}}
\newcommand{\listenfrac}[1]{\alpha_{#1}}
\newcommand{\listenfracvec}{\bm{\upalpha}}

\newcommand{\xmitcost}[1]{X_{#1}}
\newcommand{\xmitcostvec}{\mathbf{X}}
\newcommand{\xmitfrac}[1]{\beta_{#1}}
\newcommand{\xmitfracvec}{\bm{\upbeta}}

\newcommand{\indionexmit}[1]{\nu_{#1}}
\newcommand{\indionexmitest}[1]{\hat{\nu}_{#1}}
\newcommand{\numlisten}{c}
\newcommand{\numlistenest}{\hat{\numlisten}}
\newcommand{\indisomelisten}{\gamma}
\newcommand{\indisomelistenest}{\hat{\gamma}}

\newcommand{\tput}{{\cal T}}
\newcommand{\aput}{\tput_{a}}
\newcommand{\gput}{\tput_{g}}

\newcommand{\tputsigma}{\tput^{\sigma}} 
\newcommand{\tputsigmasim}{\widetilde{\tputsigma}}
\newcommand{\gputsigma}{\gput^{\sigma}} 
\newcommand{\gputsigmasim}{\widetilde{\gputsigma}} 
\newcommand{\aputsigma}{\aput^{\sigma}} 
\newcommand{\aputsigmasim}{\widetilde{\aputsigma}} 
\newcommand{\gputsigmaexp}{\widetilde{\gputsigma}} 

\newcommand{\tputconvth}{\alpha} 

\newcommand{\UB}{{\cal T}^*}
\newcommand{\UBG}{{\cal T}_{g}^{*}} 
\newcommand{\UBA}{{\cal T}_{a}^{*}} 
\newcommand{\UBNC}{{\cal T}_{\rm nc}^{*}} 
\newcommand{\UBsigma}{{\cal T}^{\oursigma}} 
\newcommand{\UBsigmaval}[1]{{\cal T}^{\oursigma = #1}} 

\newcommand{\slotlength}{\theta}
\newcommand{\periodsize}{P}
\newcommand{\optsol}[1]{{(\listenfrac{#1}^{*},\xmitfrac{#1}^{*})}}
\newcommand{\optsolvec}{(\listenfracvec^{*}, \xmitfracvec^{*})}

\newcommand{\ijfrac}{\chi_{i,j}}

\newcommand{\state}{w}
\newcommand{\numstate}{W}
\newcommand{\statevec}{\mathbf{w}}
\newcommand{\statespace}{{\mathcal{W}}}
\newcommand{\steadystate}{\pi}
\newcommand{\steadystatevec}{\bm{\uppi}}

\newcommand{\trans}{\lambda}
\newcommand{\transij}{\trans_{uv}}
\newcommand{\transSL}{\trans_{sl}}
\newcommand{\transLX}{\trans_{lx}}
\newcommand{\transLS}{\trans_{ls}}
\newcommand{\transXL}{\trans_{xl}}

\newcommand{\oursigma}{\sigma}
\newcommand{\allclear}{A}

\newcommand{\ouruW}[1]{\SI{#1}{\micro\watt}}
\newcommand{\ourmW}[1]{\SI{#1}{\milli\watt}}
\newcommand{\ourW}[1]{\SI{#1}{\watt}}
\newcommand{\ourWatt}[1]{\SI{#1}{Watt}}
\newcommand{\ourJ}[1]{\SI{#1}{\joule}}
\newcommand{\ourJoule}[1]{\SI{#1}{Joule}}

\newcommand{\ourms}[1]{\SI{#1}{\milli\second}}
\newcommand{\ours}[1]{\SI{#1}{second}}
\newcommand{\ourmF}[1]{\SI{#1}{\milli\farad}}
\newcommand{\ourF}[1]{\SI{#1}{\farad}}
\newcommand{\ourV}[1]{\SI{#1}{\volt}}
\newcommand{\ourhour}[1]{\SI{#1}{\hour}}
\newcommand{\ourminute}[1]{\SI{#1}{\minute}}
\newcommand{\oursecond}[1]{\SI{#1}{\second}}

\newcommand{\epoch}{\tau}
\newcommand{\epochk}{\epoch{k}}
\newcommand{\epochm}{\epoch{k-1}}
\newcommand{\epochp}{\epoch{k+1}}
\newcommand{\mplier}{\eta}
\newcommand{\mpliernode}[1]{\mplier_{#1}}
\newcommand{\lmnode}[1]{\mplier_{#1}} 
\newcommand{\mpliervec}{\bm{\upeta}}
\newcommand{\mplieropt}{\mplier^*}

\newcommand{\bat}{b}
\newcommand{\batnode}[1]{\bat_{#1}}
\newcommand{\stepsize}{\delta}
\newcommand{\empharvestrate}[1]{\hat{\lambda}_{#1}} 
\newcommand{\consumerate}[1]{\mu_{#1}} 
\newcommand{\empconsumerate}[1]{\hat{\mu}_{#1}} 
\newcommand{\empxmitfrac}[1]{\hat{\xmitfrac{}}_{#1}}
\newcommand{\emplistenfrac}[1]{\hat{\listenfrac{}}_{#1}}
\newcommand{\wakeuprate}[1]{\exp \left(- \frac{1}{\sigma} \lm{#1}(k) \listencost{#1} \right)}
\newcommand{\listentoxmitrate}[1]{\exp \left(\frac{1}{\sigma} \lm{#1}(k) (\listencost{#1} - \xmitcost{#1}) \right)}

\newcommand{\capsize}{C_{\rm cap}}
\newcommand{\timedilationfactor}{\chi}
\newcommand{\tputexp}{U_{\oursigma}^{\rm exp}}
\newcommand{\tputexpval}[1]{U_{\oursigma = #1}^{\rm exp}} 
\newcommand{\avgconsumerateexp}{\overline{P}} 
\newcommand{\timeintervalexp}{\Delta t} 

\newcommand{\networkprob}{\zeta}
\newcommand{\networkprobvec}{\bm{\upzeta}}
\newcommand{\gradient}{g}
\newcommand{\gradientgap}{\Delta}
\newcommand{\empgradient}[1]{\hat{g}_{#1}}
\newcommand{\onenorm}[1]{|#1|} 

\newcommand{\powerharv}{\rho}
\newcommand{\powerharvemp}{\hat{\rho}}
\newcommand{\powerharvempvec}{\hat{\uprho}}

\newcommand{\powercons}{\mu}
\newcommand{\powerconsemp}{\hat{\mu}}
\newcommand{\powerconsempvec}{\hat{\upmu}}
\newcommand{\listenfracemp}[1]{\hat{\listenfrac{}}_{#1}}
\newcommand{\xmitfracemp}[1]{\hat{\xmitfrac{}}_{#1}}

\newcommand{\markovhist}{\mathcal{F}}
\newcommand{\gradientemp}{\hat{g}}
\newcommand{\gradientempvec}{\hat{\textbf{g}}}
\newcommand{\gradientvec}{\mathbf{g}}
\newcommand{\gradientmax}{g_{\rm max}}
\newcommand{\costmax}{\bar{C}}
\newcommand{\mpliermax}{\eta_{\rm max}}
\newcommand{\expectation}[1]{\mathbb{E}\left[ #1 \right]}

\newcommand{\prob}{\mathbb{P}}
\newcommand{\slev}{\theta_{2}}
\newcommand{\norm}[1]{|| #1 ||}
\newcommand{\normtv}[1]{|| #1 ||_{\rm TV}}


\begin{abstract}
Wireless object tracking applications are gaining popularity and will soon utilize emerging ultra-low-power device-to-device communication. However, severe energy constraints require much more careful accounting of energy usage than what prior art provides. In particular, the available energy, the differing power consumption levels for listening, receiving, and transmitting, as well as the limited control bandwidth must all be considered. Therefore, we formulate the problem of maximizing the throughput among a set of heterogeneous broadcasting nodes with differing power consumption levels, each subject to a strict ultra-low-power budget.
We obtain the oracle throughput (i.e., maximum throughput achieved by an oracle) and use Lagrangian methods to design {\name} -- a simple asynchronous distributed protocol in which nodes transition between sleep, listen, and transmit states, and dynamically change the transition rates.
{\name} can operate in groupput or anyput modes to respectively maximize two alternative throughput measures.
We show that {\name} approaches the oracle throughput.
The performance is also evaluated numerically and via extensive simulations and it is shown that {\name} outperforms prior art by $6$x -- $17$x under realistic assumptions. Moreover, we evaluate {\name}'s latency performance and consider design tradeoffs when operating in groupput and anyput modes. Finally, we implement {\name} using the TI eZ430-RF2500-SEH energy harvesting nodes and experimentally show that in realistic environments it obtains $57\%$ -- $77\%$ of the achievable throughput.


\end{abstract}


\section{Introduction}

Object tracking and monitoring applications are gaining popularity within the realm of Internet-of-Things~\cite{Atzori20102787_IoT}. One enabler of such applications is the growing class of ultra-low-power wireless nodes. An example is active tags that can be attached to physical objects, harvest energy from ambient sources, and communicate tag-to-tag toward gateways \cite{Gorlatova_Mobicom, Margolies_EnHANTS_TOSN}.
Relying on node-to-node communications will require less infrastructure than traditional (RFID/reader-based) implementations. Therefore, as discussed in~\cite{Gorlatova_Mobicom, Margolies_EnHANTS_TOSN, buettner2011dewdrop, Wang_SigComm2013}, it is envisioned that such ultra-low-power nodes will facilitate tracking applications in healthcare, smart building, assisted living, manufacturing, supply chain management, and intelligent transportation.

A fundamental challenge in networks of ultra-low-power nodes is to schedule the nodes' sleep, listen/receive, and transmit events without coordination, such that they communicate effectively while adhering to their strict power budgets.
For example, energy harvesting tags need to rely on the power that can be harvested from sources such as indoor-light or kinetic energy, which provide $0.01-\ourmW{0.1}$ \cite{Gorlatova_NetworkingLowPower,gorlatova2015movers} (for more details see the review in~\cite{Ulukus_JSACreview} and references therein). These power budgets are much lower than the power consumption levels of current low-power wireless technologies such as Bluetooth Low Energy (BLE)~\cite{ble} and ZigBee/802.15.4 \cite{lee2007comparative} (usually at the order of $1-\ourmW{10}$). On the other hand, BLE and ZigBee are designed to support data rates (up to a few $\SI{}{\Mbps}$) that are higher than required by the applications our work envisages supporting (less than a few $\SI{}{\Kbps}$).

In this paper, we formulate the problem of \textit{maximizing broadcast throughput among energy-constrained nodes}. We design, analyze, and evaluate {\name}: \textbf{E}nergy-\textbf{con}strained Broad\textbf{Cast}. {\name} is an asynchronous distributed protocol in which nodes transition between sleep, listen/receive, and transmit states, while maintaining a power budget.
The nodes and network we focus on have the following characteristics:

\noindent{\bf Broadcast}:
A transmission can be heard by all listening nodes in range.

\noindent{\bf Severe power constraints}:
The power budget is so limited that each node needs to spend most of its time in sleep state and the supported data rates can be of a few $\SI{}{\Kbps}$~\cite{Gorlatova_NetworkingLowPower}. Traditional approaches that spend energy in order to improve coordination (e.g., accurate clocks, slotting, synchronization) or form some sort of structure (e.g., routing tables and clusters) are too expensive given limited energy and bandwidth.

\noindent{\bf Unacquainted}:
Nodes do not require pre-existing knowledge of their environment (e.g., properties of neighboring nodes). This can result from the restricted power budget or from unanticipated environment changes due to altered energy sources and/or node mobility.

\noindent{\bf Heterogeneous}:
The power budgets and the power consumption levels can differ among the nodes.

Efficiently operating such structureless and ultra-low-power networks requires nodes to make their sleep, listen, or transmit decisions in a distributed manner. Therefore, we consider the fundamental problem of maximizing the rate at which the messages can be delivered (the actual content of the transmitted messages depends on the application). Namely, we focus on maximizing the broadcast throughput and consider two alternative definitions: 

\begin{itemize}[topsep=0pt]
\item
{\bf Groupput} -- the total rate of successful bit transmissions to \textit{all the receivers} over time.
Groupput directly applies to tracking applications in which nodes utilize a neighbor discovery protocol to identify neighbors which are within wireless communication range~\cite{sun2014hello, Gerla14, PANDA}. In such applications, broadcasting information to all other nodes in the network is important, allowing the nodes to transfer data more efficiently under the available power budgets. Groupput can also be applied to data flooding applications where the data needs to be collected at all the nodes in a network.

\item
{\bf Anyput} -- the total rate of successful bit transmissions to \textit{at least one receiver} over time. It applies to delay-tolerant environments that utilize gossip-style methods to disseminate information. In traditional gossip communication, a node selects a communication partner in a deterministic or randomized manner. Then, it determines the content of the message to be sent  based on a naive store-and-forward, compressive sensing~\cite{luo2009compressive, srisooksai2012practical}, or decentralized
coding \cite{dimakis2005ubiquitous}. As another example, in delay-tolerant applications, data transmission may get disrupted or lost due to the limits of wireless radio range, sparsity of mobile nodes, or limited energy resources, a node may wish to send its data to any available receiver.
\end{itemize}

First, we derive oracle groupput and anyput (i.e., maximum throughput achieved by an oracle) and provide methods to efficiently compute their values. 
Then, we use Lagrangian methods and a Q-CSMA (Queue-based Carrier Sense Multiple Access) approach to design {\name}.
{\name} can operate in groupput or anyput modes to respectively maximize the two alternative throughput measures.
Nodes running {\name} dynamically adapt their transition rates between sleep, listen, and transmit states based on (i) the energy available at the node and (ii) the number (or existence) of other active listeners. To support the latter, a listening node emits a low-cost informationless ``ping'' which can be picked up by other listening nodes, allowing them to estimate the number (or existence) of active listeners. We briefly discuss how this method helps increasing the throughput and the implementation aspects. We analyze the performance of {\name} and prove that, in theory, it converges to the oracle throughput.

We evaluate the throughput performance of {\name} numerically and via extensive simulations under a wide range of power budgets and listen/transmit consumption levels, and for various heterogeneous and homogeneous networks. Specifically, numerical results show that {\name} outperforms prior art (Panda~\cite{PANDA}, Birthday~\cite{McGlynn_mobihoc01}, and Searchlight~\cite{Bakht_mobicom2012}) by a factor of $6$x -- $17$x under realistic assumptions. In addition, we consider the performance of {\name} in terms of burstiness and latency. We also consider the design tradeoffs of {\name} when oprating in groupput and anyput modes.

We implement {\name} using the TI eZ430-RF2500-SEH energy harvesting nodes and experimentally show that in practice it obtains $57\%$ -- $77\%$ of the achievable throughput. Moreover, we compare the experimental throughput to analytical throughput of Panda~\cite{PANDA} (where the analytical throughput is usually better than the experimental performance) and show that, for example,  {\name} outperforms Panda by $8$x -- $11$x.

We note that {\name} is not designed based on specific assumption, regarding the topology and that nodes do not need to know the properties of other nodes. Yet, in this paper, we mainly focus on a clique topology (i.e., all nodes are within the communication range of each other), since it lends itself to analysis. We briefly extend the analytical results to non-clique topologies and also evaluate the performance of {\name} in such networks. 

To summarize, the main contributions of this paper are:  (i) a distributed asynchronous protocol for a heterogeneous collection of energy-constrained wireless nodes, that can obtain throughput that approaches the maximum possible, (ii) efficient methods to compute the oracle throughput, and (iii) extensive performance evaluation of the protocol.

The rest of the paper is organized as follows. We discuss related work in Section~\ref{sec:vision} and formulate the problem in Section~\ref{sec:formulation}.
In Section~\ref{sec:upper}, we present methods to efficiently compute the oracle throughput.
We present {\name} in Section~\ref{sec:distributed} and the proof of the main theoretical result in Section~\ref{sec:analysis}. In Section~\ref{sec:sim}, we evaluate {\name} numerically and via simulations. We then discuss the experimental implementation and evaluation of {\name} in Section~\ref{sec:impl}. We conclude in Section~\ref{sec:conclusion}.


\section{Related Work}
\label{sec:vision}


There is vast amount of related literature on sensor networking and neighbor discovery that tries to limit energy consumption.  
Most of the protocols do not explicitly account for different listen and transmit power consumption levels of the nodes, or do not account for different power budgets~\cite{Sun_NDSurvey, ye2004medium, Gerla14, Dutta_disco08, wisemac, Bakht_mobicom2012, Kandhalu_uconnect10, sun2014hello, McGlynn_mobihoc01, vasudevan_ton13}. They mostly use a duty cycle during which nodes sleep to conserve energy and when nodes are simultaneously awake, a pre-determined listen-transmit sequence with an unalterable power consumption level is used. However, for ultra-low-power nodes constrained by severe power budgets, the appropriate amount of time a node sleeps should explicitly depend on the relative listen and transmit power consumption levels. These prior approaches achieve throughput levels which are much below optimal (and hence much below what {\name} can achieve). Additionally, these protocols often require some explicit coordination (e.g., slotting~\cite{sun2014hello, McGlynn_mobihoc01} or exchange of parameters~\cite{Bakht_mobicom2012, PANDA}), which are not suitable for emerging ultra-low-power nodes.

From the theoretical point of view, our approach is inspired by the prior work on network utility maximization (e.g.,~\cite{chiang2007layering}), and queue-based CSMA literature (e.g.,~\cite{GS10, GBW13, ME08, JW10, xu2010self}). 
However, the problem considered in this paper is not a simple extension of the prior work for two reasons. First, in the past work on CSMA and network utility maximization, nodes or links make decisions based on the relative sizes of queues. Often, a queue is a backlog of data to send or the available energy. Prior work that considers the latter (e.g., \cite{chen2014simple}) uses the energy only for transmission, while listening is ``free'', which is a very different paradigm than the one considered in this paper. Second, in our setting, the queue ``backlogs'' energy but there is no clear mapping as previously assumed from energy to successful transmission. A node's listen or transmit events will relieve the backlog, but \textit{do not} increase utility (throughput) unless other nodes are appropriately configured (i.e., transmitting when no listening nodes exist or listening when no transmitting nodes exist does not increase the throughput). This coordination of state among nodes to utilize their energy makes the considered problem more challenging.

Finally, we note that our approach should be amenable to emerging physical layer technology such as backscatter~\cite{Wang_SigComm2013}. 


\section{Model and Problem Formulation}
\label{sec:formulation}

\begin{table}[!t]
\caption{Nomenclature}
\label{table:notation}
\vspace{-0.5\baselineskip}
\footnotesize
\begin{center}
\begin{tabular}{| p{0.1\columnwidth} | p{0.8\columnwidth} |}
\hline
$\setnodes$, $\numnode$ & Set of nodes, number of nodes \\
$\listencost{i}$, $\listencostvec$ & Node $i$'s listen power consumption ($\ourWatt{}$), $\listencostvec = [\listencost{i}]$ \\
$\xmitcost{i}$, $\xmitcostvec$ & Node $i$'s transmit power consumption ($\ourWatt{}$), $\xmitcostvec = [\xmitcost{i}]$ \\
$\energybound{i}$, $\energyboundvec$ & Node $i$'s power budget ($\ourWatt{}$), $\energyboundvec = [\energybound{i}]$ \\
$\batnode{i}$ & Energy storage level of node $i$ ($\ourJoule{}$)\\
$\statevec$, $\statespace$ & Network state, the set of collision-free states \\
$\listenfrac{i}$, $\listenfracvec$ & Fraction of time node $i$ listens, $\listenfracvec = [\listenfrac{i}]$ \\
$\xmitfrac{i}$, $\xmitfracvec$ & Fraction of time node $i$ transmits, $\xmitfracvec = [\xmitfrac{i}]$ \\
$\indisomelisten$, $\indisomelistenest$ & Indicator if existing some nodes listening, its estimated value \\
$\numlisten$, $\numlistenest$ & Number of nodes listening, its estimated value \\
$\indionexmit{}$ & Indicator if there is exactly one node transmitting \\
$\steadystate_{\statevec}$, $\steadystatevec$ & Fraction of time the network is in $\statevec \in \statespace$, $\steadystatevec = [\steadystate_{\statevec}]$ \\
$\tput_{\statevec}$ & Throughput of state $\statevec \in \statespace$ \\
$\tput$, $\UB$ & Throughput and oracle throughput \\
$\gput$, $\UBG$ & Groupput and oracle groupput \\
$\aput$, $\UBA$ & Anyput and oracle anyput \\
$\mpliernode{i}$, $\mpliervec$ & Lagrange multiplier of node $i$, $\mpliervec = [\mpliernode{i}]$ \\
\hline
\end{tabular}
\end{center}
\vspace{-1\baselineskip}
\end{table}



We consider a network of $\numnode$ energy-constrained nodes whose objective is to distributedly maximize the broadcast throughput among them. The set of nodes is denoted by $\setnodes$. Table~\ref{table:notation} summarizes the notations.

\vspace{-0.5\baselineskip}
\subsection{Basic Node Model}

\noindent{\bf Power consumption}:
A node $i \in \setnodes$ can be in one of three states: \textit{sleep} ($s$), \textit{listen/receive}\footnote{We refer the \textit{listen} and \textit{receive} states synonymously as the power consumption in both states is similar.} ($l$), and \textit{transmit} ($x$), and the respective power consumption levels are $0$, $\listencost{i}~(\ourW{})$, and $\xmitcost{i}~(\ourW{})$.\footnote{The actual power consumption in \textit{sleep} state, which may be non-zero, can be incorporated by reducing $\energybound{i}$, or increasing both $\listencost{i}$ and $\xmitcost{i}$, by the sleep power consumption level.} These values are based on hardware characteristics.

\noindent{\bf Power budget}:
Each node $i$ has a \textit{power budget} of $\energybound{i}~(\ourW{})$. This budget can be the rate at which energy is harvested by an energy harvesting node or a limit on the energy spending rate such that the node can maintain a certain lifetime. In practice, the power budget may vary with time~\cite{Gorlatova_NetworkingLowPower, gorlatova2015movers} and the distributed protocol should be able to adapt.
For simplicity, we assume that the power budget is constant with respect to time. However, the analysis can be easily extended to the case with time-varying power budget with the same constant mean.
Each node $i$ also has an energy storage (e.g., a battery or a capacitor) whose level at time $t$ is denoted by $\batnode{i}(t)$.

\noindent{\bf Severe Power Constraints}:
Intermittently connected energy-constrained nodes cannot rely on complicated synchronization or structured routing approaches.

\noindent{\bf Unacquainted}:
Low bandwidth implies that each node $i$ must operate with very limited (i.e., no) knowledge regarding its neighbors, and hence, does not know or use the information $(\energybound{j}, \listencost{j}, \xmitcost{j})$ of the other nodes $j \ne i$.

\vspace{-0.1\baselineskip}
\subsection{Architecture Assumptions}
We assume that there is only \textit{one frequency channel} and \textit{a single transmission rate} is used by all nodes in the transmit state.
Similar to CSMA, nodes perform \textit{carrier sensing} prior to attempting transmission to check the availability of the medium. Energy-constrained nodes can only be awake for very short periods, and therefore, the likelihood of overlapping transmissions is negligible.
We also assume that \textit{a node in the listen state can send out low-cost, informationless ``pings''} which can be picked up by other listening nodes, allowing them to estimate the number (or existence) of active listeners. We explain in Section~\ref{sec:distributed} how this property will help us develop a distributed protocol and in Section~\ref{sec:impl}, we provide practical means by which such estimates can be obtained.

\vspace{-0.1\baselineskip}
\subsection{Model Simplifications}
At any time $t$, the network state can be described as a vector $\statevec(t) = [\state_i(t)]$, where $\state_i(t) \in \{s,l,x\}$ represents the state of node $i$.
While the distributed protocol {\name} (described in Section~\ref{sec:distributed}) can operate in general scenarios, for analytical tractability, we make the following assumptions:
\begin{itemize}[topsep=0pt]
\item The network is a \textit{clique}.\footnote{We also investigate non-clique networks in Section~\ref{ssec:upper-additional}.}
\item Nodes can perform \textit{perfect carrier sensing} in which the propagation delay is assumed to be zero.
\end{itemize}
These assumptions are suitable in the envisioned applications where the distances between nodes are small. Under these assumptions, the network states can be restricted to the set of \textit{collision-free} states, denoted by $\statespace$ (i.e., states in which there is \textit{at most} one node in transmit state).
This reduces the size of the state space from $3^N$ to $(\numnode+2) 2^{\numnode-1}$.

Let $\indisomelisten_{\statevec} \in \{0,1\}$ indicate whether there \textit{exists} some nodes listening in state $\statevec$ and let $\numlisten_{\statevec}$ be the number of listeners in state $\statevec$. We use $\indionexmit{\statevec} \in \{0,1\}$ as an indicator which is equal to $1$ if there is \textit{exactly} one transmitter in state $\statevec$ and is $0$ otherwise. Based on these indicator functions, two measures of broadcast throughput, \textit{groupput} and \textit{anyput}, and the throughput of a given network state $\statevec$ are defined below.
\begin{definition}[Groupput]
The groupput, denoted by $\gput$, is the aggregate throughput of the transmissions received by \textit{all} the receivers, where each transmitted bit is counted once per receiver to which it is delivered, i.e.,
\begin{align}
\label{eqn:maximize-varying}
\gput & = \lim_{T \to \infty} \frac{1}{T} \int_{t=0}^T \indionexmit{\statevec(t)} \numlisten_{\statevec(t)} \textrm{d}t.
\end{align}
\end{definition}
\begin{definition}[Anyput]
The anyput, denoted by $\aput$, is the aggregate throughput of the transmissions that are received by \textit{at least} one receiver, i.e.,
\begin{align}
\label{eqn:max-sender-varying}
\aput & = \lim_{T \to \infty} \frac{1}{T} \int_{t=0}^T \indionexmit{\statevec(t)} \indisomelisten_{\statevec(t)} \textrm{d}t.
\end{align}
\end{definition}
\begin{definition}[Network State Throughput]
The throughput associated with a given network state $\statevec \in \statespace$, denoted by $\tput_{\statevec}$, is defined as
\begin{equation}
\label{eq:tauw}
\tput_{\statevec} = \left\{
\begin{aligned}
& \indionexmit{\statevec} \numlisten_{\statevec}, \quad\textrm{for Groupput}, \\
& \indionexmit{\statevec} \indisomelisten_{\statevec}, \quad\textrm{for Anyput}.
\end{aligned} \right.
\end{equation}
\end{definition}

Note that without energy constraints, the oracle (maximum) groupput is $(\numnode-1)$ and is achieved when some node always transmits and the remaining $(\numnode-1)$ nodes always listen and receive the transmission. Similarly, the oracle (maximum) anyput without energy constraints is $1$ and is achieved when some node always transmits and some other node always listens and receives the transmission.

\subsection{Problem Formulation}

Define $\steadystate_{\mathbf{z}}$ as the fraction of time the network spends in a given state $\mathbf{z} \in \statespace$, i.e.,
\begin{equation}
\label{eqn:steadystate}
\steadystate_{\mathbf{z}} = \lim_{T\to\infty} \frac{1}{T} \int_{t=0}^T \mathbf{1}_{ \{\statevec(t) = \mathbf{z}\} } \, \textrm{d}t,
\end{equation}
where $\mathbf{1}_{ \{\statevec(t) = \mathbf{z}\} }$ is the indicator function which is $1$, if the network is with state $\mathbf{z}$ at time $t$, and is $0$ otherwise. Correspondingly, denote $\steadystatevec = [\steadystate_{\statevec}]$.

Below, we define the \textit{energy-constrained throughput maximization problem} {\namelp} where the fractions of time each node spends in sleep, listen, and transmit states are assigned while the node maintains the power budget. Define variables $\listenfrac{i}, \xmitfrac{i} \in [0,1]$ as the fraction of time node $i$ spends in listen and transmit states, respectively. The fraction of time it spends in sleep state is simply $(1-\listenfrac{i}-\xmitfrac{i})$. In view of~\eqref{eqn:maximize-varying} --~\eqref{eqn:steadystate}, {\namelp} is given by
\begin{eqnarray}
\label{eqn:tput-markov}
\hspace{-0.25in} {\namelp} & \hspace{-0.05in} \max\limits_{\steadystatevec}~ & \hspace{-0.1in} \sum\nolimits_{\statevec \in \statespace} \steadystate_{\statevec} \tput_{\statevec} \\
\label{eqn:energy-constraint2}
& \hspace{-0.45in} \textrm{subject to}~ & \hspace{-0.1in} \listenfrac{i} \listencost{i} + \xmitfrac{i} \xmitcost{i} \leq \energybound{i}, ~\forall i \in \setnodes, \\
\label{eqn:compute-frac2}
&& \hspace{-0.1in} \listenfrac{i} = \sum\nolimits_{\statevec \in \statespace^l_i} \steadystate_{\statevec},
~\xmitfrac{i} = \sum\nolimits_{\statevec \in \statespace^x_i} \steadystate_{\statevec}, \\
\label{eqn:pisum}
&& \hspace{-0.1in} \sum\nolimits_{\statevec \in \statespace} \steadystate_{\statevec} = 1, ~\steadystate_{\statevec} \geq0, ~\forall \statevec \in \statespace,
\end{eqnarray}
where $\statespace_i^l$ and $\statespace_i^x$ are the sets of states $\statevec \in \statespace$ in which $\state_i = l$ and $\state_i = x$, respectively. Each node is constrained by a power budget, as described in~(\ref{eqn:energy-constraint2}), and (\ref{eqn:pisum}) represents the fact that at any time, the network operates in one of the collision-free states $\statevec \in \statespace$.

Based on the solution to {\namelp}, the maximum throughput is achievable by an \textit{oracle} that can schedule nodes' sleep, listen, and transmit periods, in a centralized manner. Therefore, we define the maximum value obtained by solving {\namelp} as the \textit{oracle throughput}, denoted by $\UB$. Respectively, we define the \textit{oracle groupput} and \textit{oracle anyput} as $\UBG$ and $\UBA$.

To evaluate {\name}, it is essential to compare its performance to the oracle throughput. However, {\namelp} is a Linear Program (LP) over an exponentially large number of variables (i.e., $|\statespace|$ is exponential in $\numnode$) and is computationally expensive to solve. In Section~\ref{sec:upper}, we show how to convert {\namelp} to another optimization problem with only a linear number of variables. Note that the solution to {\namelp} only provides the optimal fraction of time each node should spend in sleep, listen, and transmit states, but \textit{does not} indicate how the nodes can make their individual sleep, listen, and transmit decisions locally. Therefore, in Section~\ref{sec:distributed}, we focus on the design of {\name} that makes these decisions based on {\namelp}.

\section{Oracle Throughput}
\label{sec:upper}

In this section, we present an equivalent LP formulation for {\namelp} in a clique network which only has a linear number of variables. We also derive both an upper and a lower bound for the oracle groupput in non-clique topologies which will be used later for evaluating the performance of {\name} in non-clique topologies.

Recall that $\listenfrac{i}$ and $\xmitfrac{i}$ are the fraction of time node $i$ spends in listen and transmit states, respectively. We can rewrite the constraints in {\namelp} as follows
\begin{align}
\label{eqn:energy-constraint}
\listenfrac{i} \listencost{i} + \xmitfrac{i} \xmitcost{i} \leq \energybound{i}, & \quad\forall i \in \setnodes, \\
\label{eqn:time-on-bound}
\listenfrac{i} + \xmitfrac{i} \leq 1, & \quad \forall i \in \setnodes, \\
\label{eqn:non-overlap-transmit} \sum\nolimits_{i \in \setnodes} \xmitfrac{i} \leq 1. &
\end{align}
Specifically,~\eqref{eqn:energy-constraint} is the usual power constraint on each node $i \in \setnodes$, and~\eqref{eqn:time-on-bound} is due to the fact that a node can only operate in one state at any time.
We remark that energy-constrained nodes can only be awake for very small fractions of time (i.e., $\listenfrac{i} + \xmitfrac{i} \ll 1$), and therefore~\eqref{eqn:time-on-bound} may be redundant. Finally, collision-free operation in a clique network where \textit{at most} one transmitter can be present at any time imposes~\eqref{eqn:non-overlap-transmit}, which bounds the sum of the transmit fractions by $1$.

\subsection{Oracle Groupput in a Clique}
\label{ssec:upper-groupput}
To maximize the groupput (\ref{eqn:maximize-varying}), it suffices that any node only listens when there is another transmitter, since listening when no one transmits wastes energy. Namely, the fraction of time node $i$ listens cannot exceed the aggregate fraction of time all other nodes transmit, i.e.,
\begin{equation}
\label{eqn:listen-when-xmit}
\listenfrac{i} \leq \sum\nolimits_{j \neq i} \xmitfrac{j}, ~\forall i \in \setnodes.
\end{equation}
Since a node only listens when there exists \textit{exactly one} transmitter, every listen counts as a reception, and the groupput of a node (i.e., the throughput it receives from all other nodes) is simply the fraction of time it spends in listen state, $\listenfrac{i}$. Therefore, the groupput in a clique network simplifies to $\sum_{i \in \setnodes} \listenfrac{i}$. The \textit{oracle groupput}, denoted by $\UBG$, can be obtained by solving the following maximization problem
\begin{eqnarray}
\label{eqn:max-groupput}
\hspace{-0.1in}{\namelpgput}
& \UBG := \max\limits_{\listenfracvec, \xmitfracvec} & \sum\nolimits_{i \in \setnodes} \listenfrac{i} \\
& \textrm{subject to} & (\ref{eqn:energy-constraint}) - (\ref{eqn:listen-when-xmit}) \nonumber.
\end{eqnarray}
{\namelpgput} is an LP consisting of $2\numnode$ variables and $(3\numnode+1)$ constraints (i.e., solving for $\listenfracvec$ and $\xmitfracvec$ given inputs of $\numnode$, $\energyboundvec$, $\listencostvec$, and $\xmitcostvec$). On a conventional laptop running Matlab, this computation for thousands of nodes takes seconds. Moreover, we show that the oracle groupput obtained by solving {\namelpgput} is indeed \textit{achievable} by an oracle which can schedule nodes' listen and transmit periods. This result is summarized in the following lemma.
\begin{lemma}
\label{lem:lp-schedule}
The (rational-valued) solution $\optsolvec$ to {\namelpgput} can be feasibly scheduled by an oracle in a fixed-size slotted environment via a periodic schedule, (perhaps) after a one-time energy accumulation interval.
\end{lemma}
\begin{IEEEproof}
The proof can be found in Appendix~\ref{append:proof-lp-schedule}.
\end{IEEEproof}

In \textit{homogeneous} networks (i.e., $\energybound{i} = \energybound{}, ~\listencost{i} = \listencost{}, ~\xmitcost{i} = \xmitcost{}, ~\forall i \in \setnodes$) where nodes are sufficiently energy-constrained (i.e.,~\eqref{eqn:energy-constraint} dominates~\eqref{eqn:time-on-bound}), the solution to {\namelpgput} is given by\footnote{We show that in the optimal solution, the equalities hold for equations~\eqref{eqn:energy-constraint} and~\eqref{eqn:listen-when-xmit}. The details are
Appendix~\ref{append:proof-lp-homo-soln}.
}
\begin{equation*}
\label{eqn:homo-soln}
\xmitfrac{}^* = \energybound{} / (\xmitcost{} + (\numnode-1)\listencost{}), ~\listenfrac{}^* = (\numnode-1) \xmitfrac{}^*, ~\UBG = \numnode \listenfrac{}^*.
\end{equation*}

\subsection{Oracle Anyput in a Clique}
\label{ssec:upper-anyput}

The oracle anyput is obtained based on the observation that a transmission only occurs when there is \textit{at least} one listener. Define variables $\ijfrac$ as the fraction of time node $j$ receives a transmission from node $i$, for the following constraints
\begin{align}
\label{eqn:enough-listen}
\xmitfrac{i} \leq \sum\nolimits_{j \neq i} \ijfrac, & \quad\forall i \in \setnodes, \\
\label{eqn:listen-overlap}
\listenfrac{j} = \sum\nolimits_{i \neq j} \ijfrac, & \quad\forall j \in \setnodes.
\end{align}
The \textit{oracle anyput}, denoted by $\UBA$, can be obtained by solving the following maximization problem
\begin{eqnarray}
\label{eqn:max-anyput}
\hspace{-0.1in}{\namelpaput}
& \UBA := \max\limits_{\listenfracvec, \xmitfracvec} & \sum\nolimits_{i \in \setnodes} \xmitfrac{i} \\
& \textrm{subject to} & (\ref{eqn:energy-constraint})-(\ref{eqn:non-overlap-transmit}), (\ref{eqn:enough-listen}), \textrm{and}~(\ref{eqn:listen-overlap}). \nonumber
\end{eqnarray}
First, \eqref{eqn:enough-listen} ensures that when node $i$ transmits, there is always \textit{at least} one other node than can receive this transmission. Then, \eqref{eqn:listen-overlap} makes sure that in the optimal schedule, the fraction of time node $j$ listens is large enough to cover all the transmissions it receives.
Therefore, {\namelpaput} maximizes the anyput by ensuring that every transmission is received by \textit{at least} one node.

In homogeneous networks, the closed-form solution to {\namelpaput} is given by
\begin{equation*}
\xmitfrac{}^* = \listenfrac{}^* = \energybound{} / (\xmitcost{} + \listencost{}), ~\UBA = \numnode \xmitfrac{}^*.
\end{equation*}

\subsection{Oracle Groupput in Non-cliques}
\label{ssec:upper-additional}

The problem formulations {\namelp} -- {\namelpaput} so far have assumed a clique network.
Obtaining the exact maximum groupput for non-cliques (denoted by $\UBNC$) is difficult. This is because a node may receive simultaneous transmissions from two nodes which are not within communication range of each other. As explained before, listen and transmit events are rare within energy-constrained nodes. Therefore, the likelihood of simultaneous transmissions is small and it is expected to have minimal impact on the throughput.

We present both an upper bound $\overline{\UBNC}$ and a lower bound $\underline{\UBNC}$ on the maximum groupput in non-clique topologies. In the scenarios where $\overline{\UBNC}$ and $\underline{\UBNC}$ are the same, the exact maximum groupput $\UBNC$ can be obtained. The lower bound $\underline{\UBNC}$ is obtained by solving {\namelpgput} but replace constraint (\ref{eqn:listen-when-xmit}) by
\begin{equation*}
\listenfrac{i} \leq \sum\nolimits_{i \in \neighborofi} \xmitfrac{j}, ~\forall i \in \setnodes,
\end{equation*}
where $\neighborofi$ is the set of neighboring nodes of node $i$. This ensures that the fraction of time node $i$ listens cannot exceed the sum of its neighboring nodes' fractions of transmissions. The upper bound $\overline{\UBNC}$ is obtained by solving {\namelpgput} in which the constraint (\ref{eqn:non-overlap-transmit}) is removed. This allows overlapping transmissions which can possibly happen in non-cliques. Numerical results show that with certain topologies, $\overline{\UBNC} = \underline{\UBNC}$ holds, resulting in the exact maximum groupput $\UBNC$. In Section~\ref{ssec:sim-nonclique}, we compute $\UBNC$ and evaluate the performance of {\name} in non-clique topologies.


\section{Distributed Protocol}
\label{sec:distributed}

In this section, we describe {\name} from the perspective of a single node that transitions between sleep, listen, and transmit states, under a power budget. Since we focus on a single node $i$, in parts of this section, we drop the subscript $i$ of previously defined variables for notational compactness.

\subsection{A Simple Heterogeneous Example}
\label{ssec:distributed-lp-example}
To better understand the challenges faced in designing {\name}, consider a simple example of $4$ nodes, all having identical listen and transmit power consumption $\listencost{i} = \xmitcost{i} = \ourmW{1}~(i = 1,2,3,4)$, but different power budgets $\energybound{i}$, as indicated in Table~\ref{table:hetero-example}.
Table~\ref{table:hetero-example} also shows the percentage of time each node spends in listen and transmit states $\optsol{i}~(i = 1,2,3,4)$ such that the groupput is maximized by solving {\namelp}. It also shows the percentage of time each node spends in transmit state {\em when awake} (i.e., $\frac{100 \cdot \xmitfrac{i}^*}{\listenfrac{i}^* + \xmitfrac{i}^*} \%$).

\begin{table}[!t]
\caption{A simple example in a heterogeneous network.}
\label{table:hetero-example}
\vspace{-0.5\baselineskip}
\small
\begin{center}
\begin{tabular}{|c|c|c|c|c|}
\hline
Node & $1$ & $2$ & $3$ & $4$ \\
\hline
Power Budget: $\energybound{i}(\ourmW{})$ & $0.005$ & $0.01$ & $0.05$ & $0.1$ \\
\hline
$\textrm{Awake}(\%)$: $\listenfrac{i}^* + \xmitfrac{i}^*$ & $0.5$ & $1.0$ & $5.0$ & $10.0$ \\
\hline
$\textrm{Transmit when Awake}(\%)$& $20.0$ & $22$ & $53.6$ & $65.7$ \\
\hline
\end{tabular}
\vspace{-0\baselineskip}
\end{center}
\end{table}

If, instead, all nodes have the same power budget of $\energybound{i} = \ourmW{0.1}$, the percentage of time each node spends in transmit state when awake is $25\%$ (with $\listenfrac{i}^* = 0.075, \xmitfrac{i}^* = 0.025$, $i = 1,2,3,4$). Note that in the above example, the power budget of node $4$ remains unchanged but changes in other nodes' power budgets shift the percentage of time it should transmit when awake from $25\%$ to $65.7\%$. This clearly shows that the partitioning of a node's power budget among listen and transmit states is highly dependent on other nodes' properties. \textit{However, we will show that if a node does not know the properties of its neighbors, an optimal configuration can be obtained without explicitly solving {\namelp}}.


\subsection{Protocol Description}

\begin{figure}[t]
\begin{center}
\includegraphics[width=0.8\columnwidth]{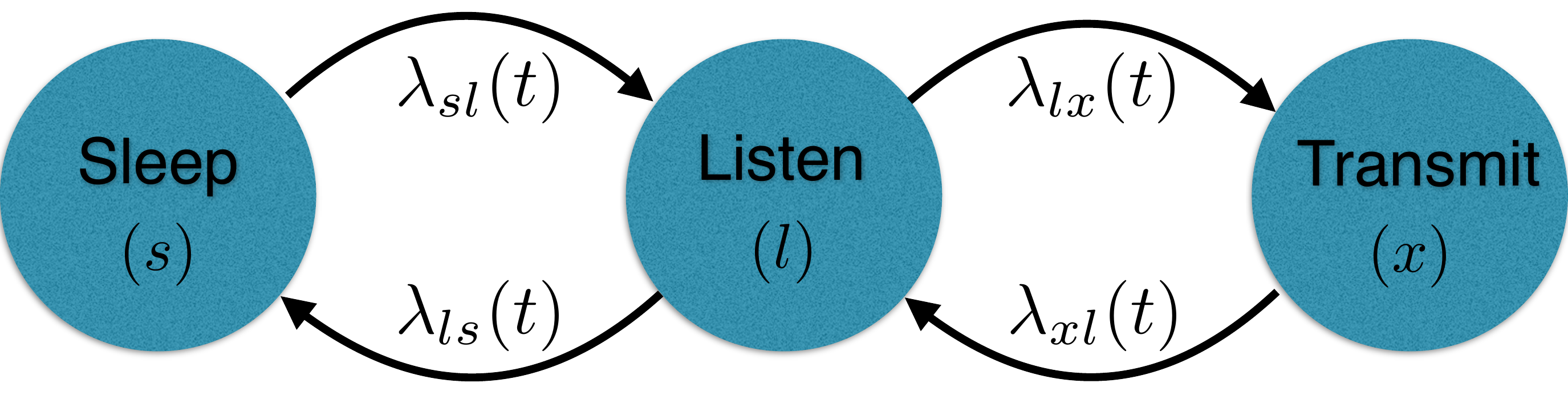}
\caption{The node's states and transition rates.}
\label{fig:3state-markov}
\vspace{-\baselineskip}
\end{center}
\end{figure}

To clearly present {\name}, we start from a theoretical framework and slowly build on it to address practicalities. As mentioned in Section~\ref{sec:formulation}, a node can be in one of three states: sleep ($s$), listen ($l$), and transmit ($x$). As depicted in Fig.~\ref{fig:3state-markov}, it must pass through listen state to transition between sleep and transmit states.
The time duration a node spends in state $u$ before transitioning to state $v$ is exponentially distributed with rate $\transij(t)$. These \textit{transition rates} can be adjusted over time. We remark that sending packets with exponentially distributed length (i.e., a node transitions from transmit state to listen state with a rate $\transXL$) is impractical. However, it can be shown that this is equivalent to continuously transmitting back-to-back unit-length packets with probability $(1-\transXL)$ if $\transXL \in [0,1]$, which is indeed the case in {\name}.

To maximize the groupput or anyput, {\name} can operate in \textit{groupput mode} or \textit{anyput mode}, respectively.
The throughput as a function of $\steadystate_{\statevec}$ (see~\eqref{eqn:tput-markov}) is controlled by appropriately adjusting the transition rates between different states of each node. {\name} determines in a distributed manner how these adjustments are performed over time. Roughly speaking, each node adjusts its transition rates $\transij(t)$ based on limited information that can be obtained in practice, including
\begin{itemize}[topsep=0pt]
\item
Its power consumption levels in listen and transmit states, $\listencost{}$ and $\xmitcost{}$, and energy storage level $\bat(t)$.
\item
A sensing of transmit activity of other nodes over the channel (CSMA-like carrier sensing).
\item
A count of other active listeners (in groupput mode), $\numlisten(t)$, or an indicator of whether there are any active listeners (in anyput mode), $\indisomelisten(t)$. In practice, $\numlisten(t)$ and $\indisomelisten(t)$ may not be accurate, and we denote $\numlistenest(t)$ and $\indisomelistenest(t)$ as their estimated values.
\end{itemize}

Note that in {\name}, unlike in previous work such as Panda~\cite{PANDA}, each node \textit{does not} need to know the number of nodes in the network, $\numnode$, and the power budgets and consumption levels of other nodes. Furthermore, a node \textit{does not} need to know its power budget $\energybound{}$ explicitly (e.g., in the case of energy harvesting~\cite{Margolies_EnHANTS_TOSN}), although this knowledge can be incorporated, if available.

Under {\name}, a node sets $\transSL(t)$ as an increasing function of the available stored energy, $\bat(t)$, to more aggressively exit sleep state. Furthermore, it sets $\transLX(t)$ as an increasing function of the number of listeners, $\numlistenest(t)$, to enter transmit state more frequently when more nodes are listening. We will describe how these functions are chosen in Section~\ref{ssec:distributed-set-trans-rate}.

\subsection{Estimating Active Listeners: Pings}
\label{sec:ping}
We now discuss the estimation of $\numlistenest(t)$ or $\indisomelistenest(t)$. Recall from Section~\ref{sec:formulation} that nodes can send out periodic \textit{pings} that any other listener can receive. The pings need not carry any explicit information and are potentially significantly cheaper and shorter than control packet transmissions (e.g., an ACK). Therefore, they consume less power and take much less time than a minimal data transmission.

Consider the case in which all nodes are required to send pings at a pre-determined rate and the power consumption is accounted for in the listening power consumption $\listencost{}$. In such a case, a fellow listener detecting such pings (e.g., using a simple energy detector) can use the count of such pings in a given period of time, or the inter-arrival times of pings, to estimate the number of active listeners $\numlisten(t)$. Estimating $\indisomelisten(t)$ is even easier by detecting the existence of any ping. In general, the estimates do not need to be accurate for {\name} to function, although poor estimates are expected to reduce throughput.

\subsection{Two Variants of {\name}}
\label{ssec:distributed-variants}

We now address the incorporation of the estimates $\numlistenest(t)$ and $\indisomelistenest(t)$ into {\name}. We present two versions of {\name} which only differ when a node is in transmit state:
\begin{itemize}[topsep=0pt]
\item
{\namecap} (the capture version): a node may ``capture'' the channel and transmit for an exponential amount of time (i.e., several back-to-back packets). When each packet transmission is completed, the transmitter listens for pings for a fixed-length \textit{pinging interval}. Each successful recipient of the transmission initiate one ping at time chosen uniformly at random on this interval. The transmitter then estimates $\numlistenest(t)$ or $\indisomelistenest(t)$ based on the count of pings received and adjusts $\transXL(t)$ (as described in Section~\ref{ssec:distributed-set-trans-rate}). In Section~\ref{ssec:impl-practical-ping}, we discuss the experimental implementation of this process.

\item
{\namenoncap} (the non-capture version): a node always releases the channel after one packet transmission. Each node continuously pings and receives pings from other nodes when listening, estimates $\numlistenest(t)$ or $\indisomelistenest(t)$, and adjusts $\transLX(t)$ (as described in Section~\ref{ssec:distributed-set-trans-rate}).
\end{itemize}
{\namecap} is significantly easier to implement since the estimates are only needed for the transmitter right after each packet transmission. The probability that the same transmitter will continue transmitting depends on the estimates $\numlistenest(t)$ or $\indisomelistenest(t)$. Therefore, our implementation and experimental evaluations in Section~\ref{sec:impl} focus on {\namecap}.

\subsection{Setting Transition Rates}
\label{ssec:distributed-set-trans-rate}
Consider a node running {\name}.
Time is broken into intervals of length $\epoch_{k}~(k=1,2,\cdots)$. The $k$-th interval is from time $t_{k-1}$ to time $t_{k}$ and we let $t_0=0$. {\name} takes input of two internal variables:
\begin{itemize}[topsep=0pt]
\item
$\mplier$ is a multiplier which is updated at the beginning of each time interval. Let $\bat[k]~(k=0,1,\cdots)$ denote the energy storage level at the end of the $k$-th time interval. Let $(\cdot)^{+}$ denote $\max(0,\cdot)$ and $\mplier[k]$ is updated as follows
\begin{equation}
\label{eqn:update-lm-epoch}
\mplier[k] = \Big(\mplier[k-1] - \stepsize_{k} / \epoch_{k} \cdot (\bat[k] - \bat[k-1]) \Big)^{+},
\end{equation}
in which $\stepsize_k\in (0,1)$ is a step size and $\bat[k] = \bat(t_{k})$.
We use square brackets here to imply that the multiplier $\mplier[k]$ remains constant for $t \in [t_{k}, t_{k+1})$.

\item
$A(t)$ is the carrier sensing indicator of a node, which is $1$ when the node does not sense any ongoing transmission, and is $0$ otherwise. Carrier sensing forces a node to ``stick'' to its current state. When receiving an ongoing transmission, a node in listen state will not exit the listen state until it finishes receiving the full transmission, and a node in sleep state will not leave the sleep state (i.e., it enters the listen state but immediately leaves when it senses the ongoing transmission).
\end{itemize}

The transition rates are described as follows (the superscripts $C$ and $N$ denote {\namecap} and {\namenoncap}), in which the two throughput modes only differ in $\transXL(t)$ (for the capture version) or in $\transLX(t)$ (for the non-capture version). For \emph{groupput} maximization, at any time $t$ in the $k$-th interval,
\begin{subequations}
\label{eqn:rates2}
\begin{align}
\label{eqn:trans-rate-sl}
\transSL(t) & = \allclear(t) \cdot \exp (-\mplier[k] \listencost{} / \oursigma), \\
\label{eqn:trans-rate-ls}
\transLS(t) & = \allclear(t), \\
\label{eqn:trans-rate-lx-cap}
\transLX^C(t) & = \allclear(t) \cdot \exp (\mplier[k] (\listencost{} - \xmitcost{}) / \oursigma), \\
\label{eqn:trans-rate-lx-noncap}
\transLX^N(t) & = \allclear(t) \cdot \exp (\mplier[k] (\listencost{} - \xmitcost{}) / \oursigma + \numlistenest(t) / \oursigma), \\
\label{eqn:trans-rate-xl-cap}
\transXL^C(t) & =  \exp (-\numlistenest(t) / \oursigma), \\
\label{eqn:trans-rate-xl-noncap}
\transXL^N(t) & =  1.
\end{align}
\end{subequations}
For \emph{anyput} maximization, $\numlistenest(t)$ is replaced with $\indisomelistenest(t)$.
Theorem~\ref{thm:convergence} below states the main result of this paper and the proof is in Section~\ref{sec:analysis}.
\begin{theorem}
\label{thm:convergence}
Let $\oursigma \to 0$ and select parameters $\stepsize_{k}$ and $\epoch_{k}$ properly (e.g., $\stepsize_{k} = 1/[(k+1)\log{(k+1)}]$ and $\epoch_{k} = k$). Under perfect knowledge of $\numlisten(t)$ or $\indisomelisten(t)$, the average throughput of {\name} ($\gput$ or $\aput$) converges to the oracle throughput ($\UBG$ or $\UBA$) given by {\namelp}.
\end{theorem}

\subsection{Stability and Choice of $ \oursigma$,  $\stepsize_k$, and $\epoch_{k}$}
\label{ssec:distributed-param-choice}
{\name} is adaptive and, as expected, it must deal with the tradeoff of ``adapting quickly but poorly'' to ``adapting optimally but slowly''.
This adaptation manifests itself into the parameters $\oursigma$, $\stepsize_{k}$, and $\epoch_{k}$.
When $\oursigma$ is decreased, the throughput increases, as we will describe in Section~\ref{sec:analysis}. However, the \emph{burstiness} also increases with respect to decreased $\oursigma$. The burstiness is a characteristic of communication involving multiple packets that are successfully received in bursts. In Section~\ref{sec:sim}, we describe how the burstiness can be analyzed and measured.

Under a given value of $\oursigma$, each node continuously adjusts the rates $\transij(t)$ based on its multiplier $\mplier$ according to~\eqref{eqn:update-lm-epoch}, which is a function of the ratio $\stepsize_{k} / \epoch_{k}$. Small $\stepsize_{k} / \epoch_{k}$ ratios make smaller changes of $\mplier$ over time, and lead to longer convergence time to the ``right'' multiplier values.
In contrast, larger $\stepsize_{k} / \epoch_{k}$ ratios make $\mplier$ oscillate more wildly near the optimal value, such that the performance of {\name} is further from the optimal. Although the guaranteed convergence requires careful choices of the parameters (as stated in Theorem~\ref{thm:convergence}), in practice, we can choose $\delta_{k} = \stepsize$ and $\epoch_{k} = \epoch$ for some small constant $\stepsize$ and large constant $\epoch$.


\section{Proof of Theorem 1}
\label{sec:analysis}

The proof of Theorem~\ref{thm:convergence} is based on a Markov Chain Monte Carlo (MCMC) approach~\cite{GBW13, JW10} from statistical physics and consists of three parts:
(i) we compute the steady state distribution of the network Markov chain under {\name} with fixed Lagrange multiplier vector $\mpliervec = [\mplier_{i}]$,
(ii) we present an alternative concave optimization problem whose optimal value approaches that of {\namelp} as $\oursigma \to 0$ and show that the steady state distribution of {\name} is indeed the optimal solution to this alternative optimization problem when the Lagrange multipliers are chosen optimally, and
(iii) we show that under {\name}, nodes update their Lagrange multipliers locally according to a ``noisy'' gradient descent which converge to the optimal Lagrange multipliers with proper choices of step sizes and interval lengths as given in Theorem~\ref{thm:convergence}.



\subsection*{Part (i): Steady State Distribution}
\label{ssec:proof-part-1}

The following lemma describes the network state distribution generated by {\name} when $\mpliervec$ freezes.
\begin{lemma}
\label{lem:dbe} 
With fixed $\mpliervec$, the network Markov chain, resulted from overall interactions among the nodes according to the transition rates (\ref{eqn:rates2}), has the steady state distribution
\begin{eqnarray}
\label{eqn:steady-state-prob}
\steadystate_{\statevec}^{\mpliervec} = \frac{1}{Z^{\mpliervec}} \exp \left[ \frac{1}{\oursigma} \left( \tput_{\statevec} - \sum\limits_{i:\state_i=l} \mpliernode{i} \listencost{i} - \sum\limits_{i:\state_i=x} \mpliernode{i} \xmitcost{i} \right) \right],
\end{eqnarray}
where $Z^{\mpliervec}$ is a normalizing constant so that $\sum\nolimits_{\statevec \in \statespace} \steadystate_{\statevec}^{\mpliervec} = 1$.
\end{lemma}
\begin{IEEEproof}
The proof can be found in Appendix~\ref{append:proof-dbe}.
\end{IEEEproof}

\subsection*{Part (ii): An Alternative Optimization}

We then present an optimization problem {\namecp} as follows
\begin{eqnarray}
\label{eqn:tput-lagrangian}
\hspace{-0.3in} {\namecp}
& \hspace{-0.1in} \max\limits_{\steadystatevec} & \sum\nolimits_{\statevec \in \statespace} \steadystate_{\statevec} \tput_{\statevec} - \oursigma \sum\nolimits_{\statevec \in \statespace} \steadystate_{\statevec} \log{\steadystate_{\statevec}}\\
& \hspace{-0.40in} \textrm{subject to} & (\ref{eqn:energy-constraint2}), ~(\ref{eqn:compute-frac2}), \textrm{and}~(\ref{eqn:pisum}) \nonumber,
\end{eqnarray}
where $\oursigma$ is the positive constant used in {\name} (the counterpart in statistical physics is the temperature in systems of interacting particles).
Note that {\namecp} is a concave maximization problem and as $\oursigma \to 0$, the optimal value of {\namecp} approaches that of {\namelp}.
To solve {\namecp}, consider the Lagrangian $\mathcal{L}(\steadystatevec,\mpliervec)$ formulated by moving the power constraint~\eqref{eqn:energy-constraint2} into the objective~\eqref{eqn:tput-lagrangian} with a Lagrange multiplier $\mpliernode{i} \geq 0$ for each node $i$, i.e.,
\begin{eqnarray}
\mathcal{L}(\steadystatevec, \mpliervec) = & \sum\nolimits_{\statevec \in \statespace} \steadystate_{\statevec} \tput_{\statevec} - \oursigma \sum\nolimits_{\statevec \in \statespace} \steadystate_{\statevec} \log{\steadystate_{\statevec}} \nonumber \\
& - \sum\nolimits_{i \in \setnodes} \left[ \mpliernode{i} (\listenfrac{i} \listencost{i} + \xmitfrac{i} \xmitcost{i} - \energybound{i}) \right].
\end{eqnarray}
In view of~\eqref{eqn:compute-frac2} and~\eqref{eqn:pisum}, it can be shown that with fixed $\mpliervec$, the optimal $\steadystatevec^{\mpliervec} = [\steadystate_{\statevec}^{\mpliervec}]$ that maximizes $\mathcal{L}(\steadystatevec, \mpliervec)$ is exactly given by~\eqref{eqn:steady-state-prob}. Therefore, if {\name} knows the optimal Lagrange multiple vector $\mpliervec^{*}$, it can start with $\mpliervec^{*}$ and the steady state distribution generated by {\name} will converge to the optimal solution to {\namecp}.

\begin{algorithm}[!t]
\small
\caption{Gradient Descent Algorithm}
\label{alg:gradient-descent}
\begin{algorithmic}[1]
\Statex\hspace{-\algorithmicindent}{\bf Input parameters}:
$\oursigma$, $\energyboundvec$, $\listencostvec$, and $\xmitcostvec$
\Statex\hspace{-\algorithmicindent}{\bf Initialization}:
$\listenfrac{i}(0)=\xmitfrac{i}(0)=\mpliernode{i}(0)=0, ~\forall i \in \setnodes$
\State {\bf for} $k=1,2,\cdots$ {\bf do}
\State ~\quad $\stepsize{}(k) = 1/k$, compute $\steadystatevec(k)$ from (\ref{eqn:steady-state-prob}) using $\mpliervec = \mpliervec(k)$
\State ~\quad {\bf for} $i=1,2,\cdots,\numnode$ {\bf do}
\State ~\quad~\quad Update $\mpliernode{i}(k)$, $\listenfrac{i}(k)$, and $\xmitfrac{i}(k)$ according to~\eqref{eqn:update-lm},~\eqref{eqn:compute-frac}
\end{algorithmic}
\normalsize
\end{algorithm}

In order to find $\mpliervec^{*}$, consider the dual function $\mathcal{D}(\mpliervec) = \mathcal{L}(\steadystatevec^{\mpliervec}, \mpliervec)$ over $\mpliervec \succeq \bm{0}$ (here $\bm{0}$ is an $\numnode$-dimensional zero vector and $\succeq$ denotes component-wise inequality). Interestingly, it can be shown that the partial derivative of $\mathcal{D}(\mpliervec)$ with respect to $\mpliernode{i}$ is simply given by
\begin{equation}
\label{eq:gradient1}
\partial \mathcal{D} / \partial\mpliernode{i} = \energybound{i} - (\listenfrac{i} \listencost{i} + \xmitfrac{i} \xmitcost{i}),
\end{equation}
which is the difference between the power budget $\energybound{i}$ and the average power consumption of node $i$. Therefore, the dual can be minimized by using a gradient descent algorithm with inputs of step size $\stepsize_k > 0$, $\energyboundvec$, $\listencostvec$, and $\xmitcostvec$, which generates a state probability $\steadystatevec(k) ~(k = 1,2,\cdots)$. This algorithm is described in Algorithm~\ref{alg:gradient-descent} along with the following equations
\begin{align}
\label{eqn:update-lm}
& \hspace{-0.1in} \mpliernode{i}(k) = \left[ \mpliernode{i}(k-1) - \stepsize_k (\energybound{i} - \listenfrac{i}(k) \listencost{i} - \xmitfrac{i}(k) \xmitcost{i}) \right]^+, \\
\label{eqn:compute-frac}
& \hspace{-0.1in} \listenfrac{i}(k) = \sum\nolimits_{\statevec \in \statespace_{i}^{l}} \steadystate_{\statevec}^{\mpliervec(k)}, ~\xmitfrac{i}(k) = \sum\nolimits_{\statevec \in \statespace_{i}^{x}} \steadystate_{\statevec}^{\mpliervec(k)}.
\end{align}
Hence, with the right choice of step size $\stepsize_k$ (e.g., $\stepsize_k=1/k$), $\steadystatevec(k)$ converges to the optimal solution to {\namecp}.

To arrive at a distributed solution, instead of computing the quantities $\listenfrac{i}$ and $\xmitfrac{i}$ directly according to (\ref{eqn:compute-frac}) (which is centralized with high complexity), {\name} approximates the difference between the power budget and the average power consumption~\eqref{eq:gradient1} by observing the dynamics of the energy storage level of each node.
Specifically, each node $i$ can update its Lagrange multiplier $\mpliernode{i}(k)$ based on the difference between its energy storage levels at the end and the start of an interval of length $\epoch_{k}$, divided by $\epoch_{k}$, as described by (\ref{eqn:update-lm-epoch}). Therefore, $\mpliernode{i}$ is updated according to a ``noisy'' gradient descent.
However, it follows from stochastic approximation (with Markov modulated noise) that by choosing step sizes and interval lengths as given in Theorem~\ref{thm:convergence}, these noisy updates will converge to $\mpliervec^*$ as $k \to \infty$ (see e.g., Theorem~1 of~\cite{jiang2009convergence}).
As mentioned in Section~\ref{ssec:distributed-param-choice}, the choice of parameters $\oursigma$, $\stepsize_{k}$, and $\epoch_{k}$ will affect the tradeoff between convergence time and the performance of {\name}.

\subsection*{Part (iii): Convergence Analysis}
The detailed proof uses similar techniques as in the proof of Theorem 1 in~\cite{jiang2009convergence} with minor modifications and can be found
in Appendix~\ref{append:proof-convergence}.


\section{Numerical Results}
\label{sec:sim}

In this section, we evaluate the throughput and latency performance of {\name} when operating in groupput and anyput modes.
%
We use the following notation: (i) $\UBG$ ($\UBA$) is the oracle groupput (anyput) obtained by solving {\namelp} or, equivalently, {\namelpgput}, (ii) $\gputsigma$ ($\aputsigma$) is the achievable groupput (anyput) of {\name} with a given value of $\oursigma$ obtained by solving {\namecp}, and (iii) $\gputsigmasim$ ($\aputsigmasim$) is the groupput (anyput) of {\name} obtained via simulations with a given value of $\oursigma$. For brevity, we ignore the subscripts of $\tputsigma$ when describing results that are general for both groupput and anyput.

\subsection{Setup}
\label{ssec:sim-setup}
We consider clique networks\footnote{We evaluate the throughput performance of {\namecaption} in non-clique topologies in~\ref{ssec:sim-nonclique}.} with $\oursigma \in \{0.1, 0.25, 0.5\}$. The nodes' power budgets and consumption levels correspond to the energy harvesting budgets and ultra-low-power transceivers in~\cite{Gorlatova_NetworkingLowPower, gorlatova2015movers, philips2014ultra}. 
Note that the performance of {\name} only depends on the \textit{ratio} between the listen or transmit power and the power budget. For example, nodes with $\energybound{} = \ouruW{10}, \listencost{} = \xmitcost{} = \ourmW{500}$ behave exactly the same as nodes with $\energybound{} = \ourmW{1}, \listencost{} = \xmitcost{} = \ourmW{50}$. Therefore, the oracle throughput applies and {\name} can operate in very general settings.

In the simulations, each node has a constant power input at the rate of its power budget, and adjusts the transition rates based on the dynamics of its energy storage level. Since nodes perform carrier sensing when waking up, there are no simultaneous transmissions and collisions. We also assume that the packet length is $\ourms{1}$ and that nodes have accurate estimate of the number of listeners or the existence of any active listeners, i.e., $\numlistenest(t) = \numlisten(t)$ or $\indisomelistenest(t) = \indisomelisten(t)$.

The simulation results show that $\tputsigmasim$ perfectly matches $\tputsigma$ for $\oursigma \in \{0.25, 0.5\}$. For $\oursigma = 0.1$, $\tputsigmasim$ does not converge to $\tputsigma$ within reasonable time due to the bursty nature of {\name}, as will be described in Section~\ref{ssec:sim-burstiness}. Therefore, we evaluate the throughput performance of {\name} by comparing $\tputsigma$ to $\UB$ with varying $\oursigma$ in both heterogeneous and homogeneous networks. Specifically, \textit{homogeneous networks} consist of nodes with the same power budget and consumption levels, i.e., $\energybound{i} = \energybound{}, \listencost{i} = \listencost{}, \xmitcost{i} = \xmitcost{},\forall i \in \setnodes$.
The simulation results also confirm that nodes running {\name} consume power on average at the rate of their power budgets.

\subsection{Heterogeneous Networks -- Throughput}
\label{ssec:sim-het-net}


\begin{figure}[!t]
\centering
\vspace{-0.75\baselineskip}
\subfloat[]{
\label{fig:sim-hetnet-gput}
\includegraphics[width=0.5\columnwidth]{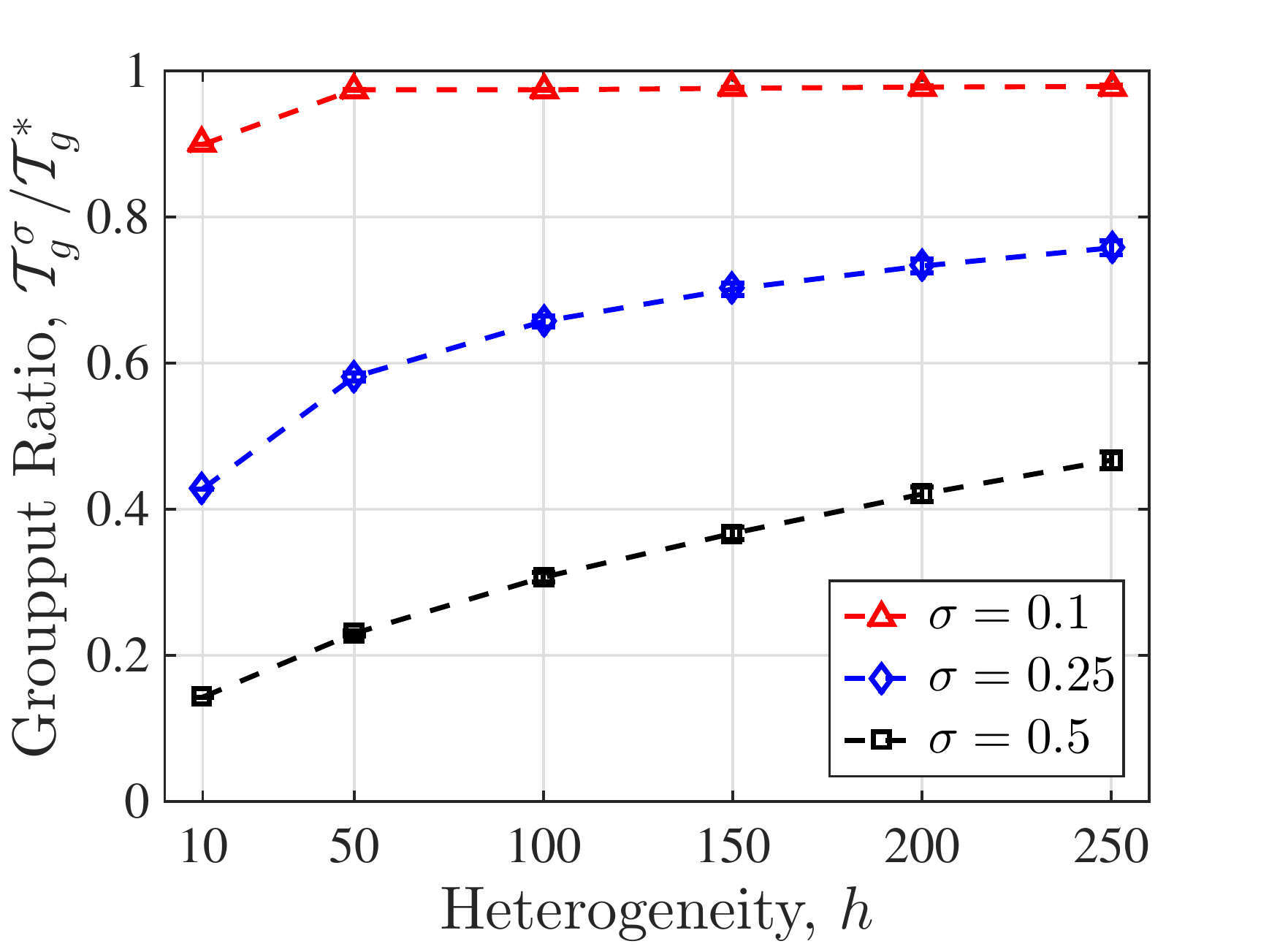}
}
\hspace{-18pt} \hfill
\subfloat[]{
\label{fig:sim-hetnet-aput}
\includegraphics[width=0.5\columnwidth]{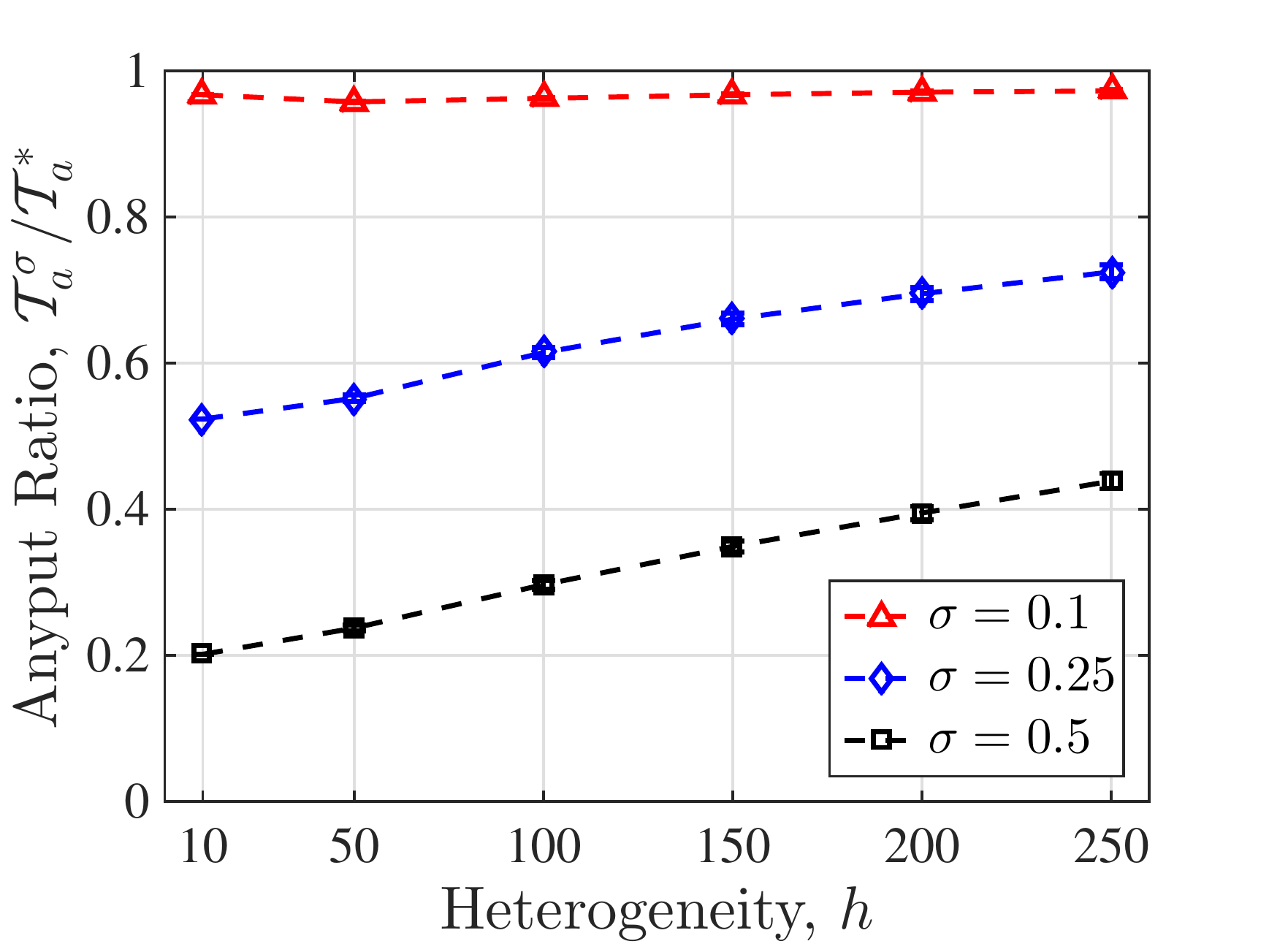}
}
\vspace{-0.5\baselineskip}
\caption{Sensitivity of the achievable throughput normalized to the oracle throughput, $\tputsigma/\UB$, for: (a) groupput and (b) anyput, to the heterogeneity of the power budget, $\energybound{}$, and power consumption levels, $\listencost{}$ and $\xmitcost{}$.}
\label{fig:sim-hetnet}
\vspace{-\baselineskip}
\end{figure}

One strength of {\name} is its ability to deal with \textit{heterogeneous} networks. Fig.~\ref{fig:sim-hetnet} shows the groupput and anyput achieved by {\name} normalized to the corresponding oracle groupput and anyput (i.e., $\tputsigma / \UB$), for heterogeneous networks with $\numnode = 5$ and $\oursigma \in \{0.1, 0.25, 0.5\}$.
Intuitively, higher values of $\tputsigma/\UB$ indicate better performance of {\name}.

Along the $x$-axis, the \textit{network heterogeneity}, denoted by $h$, is varied from $10$ to $250$ at discrete points. The relationship between the network heterogeneity and the values of $h$ is as follows: (i) for each node $i$, $\listencost{i}$ and $\xmitcost{i}$ are independently selected from a uniform distribution on the interval $[510-h, 490+h]~(\ouruW{})$, (ii) for each node $i$, a variable $h'$ is first sampled from the interval $[-\log{\frac{h}{100}}, \log{h}]$ uniformly at random, and then $\energybound{i}$ is set to be $\exp{(h')}$. Therefore, the energy budget $\energybound{i}$ varies from $100/h$ to $h~(\ouruW{})$. As a result, for any $h$, $\listencost{i}$ and $\xmitcost{i}$ have mean values of $\ouruW{500}$, and $\energybound{i}$ has median of $\ouruW{10}$ but its mean increases as $h$ increases. Note that a homogeneous network is represented by $h=10$. 

The $y$-axis indicates for each value of $h$, the mean and the $95\%$ confidence interval of the ratios $\tputsigma / \UB$ averaged over $1000$ heterogeneous network samples. Specifically, in each network sample, each node $i$ samples $(\energybound{i}, \listencost{i}, \xmitcost{i})$ according to the processes described above.
Figs.~\ref{fig:sim-hetnet}\subref{fig:sim-hetnet-gput} and~\ref{fig:sim-hetnet}\subref{fig:sim-hetnet-aput} show that the network heterogeneity with respect to both the nodes' power budgets and consumption levels increases as $h$ increases. Fig.~\ref{fig:sim-hetnet} also shows that the throughput ratio $\tputsigma / \UB$ increases as $\oursigma$ decreases, and approaches $1$ as $\oursigma \to 0$.
Furthermore, with increased heterogeneity of the network, the throughput ratio has little dependency on the network heterogeneity $h$ but heavy dependency on $\oursigma$.
In general, the groupput and anyput ratios are similar except for homogeneous networks ($h=10$). In such networks, the anyput ratio is slightly higher than the groupput ratio. This is due to the fact that nodes have the same values of $\energybound{i}$, $\listencost{i}$, and $\xmitcost{i})$. Therefore, determining the existence of any active listeners, $\indisomelisten(t)$, is easier than determining the number of active listeners, $\numlisten(t)$.



\subsection{Homogeneous Networks -- Throughput and Comparison to Related Work }
\label{ssec:sim-tput}

We now evaluate the throughput of {\name} in \textit{homogeneous} networks with $\numnode = 5$, $\energybound{} = \ouruW{10}$, $\listencost{} + \xmitcost{} = \ourmW{1}$, and $\oursigma \in \{0.1, 0.25, 0.5\}$. We also compare the groupput achieved by {\name} to three protocols in related work: \textit{Panda}~\cite{PANDA}, \textit{Birthday}~\cite{McGlynn_mobihoc01}, and \textit{Searchlight}~\cite{Bakht_mobicom2012}, which operate under stricter assumptions than {\name}. In particular:
\begin{itemize}
\item
The probabilistic protocols Panda and Birthday both require a homogeneous set of nodes and a priori knowledge of $\numnode$. The throughput of Panda and Birthday is computed as described in~\cite{PANDA} and~\cite{McGlynn_mobihoc01}, respectively.
\item
The deterministic protocol Searchlight is designed for minimizing the worst case pairwise discovery latency, which does not directly address multi-party communication across a shared medium. However, the discovery latency is closely related to the throughput, since the inverse of the average latency is the throughput. Hence, maximizing throughput is equivalent to minimizing the average discovery latency. We derive an upper bound on the throughput of Searchlight by multiplying the pairwise throughput by $(\numnode-1)$. This is assuming that all other $(\numnode-1)$ nodes will be receiving when one node transmits. However, in practice the throughput is likely to be lower unless all the nodes are synchronized and coordinated.
\end{itemize}


\begin{figure}[!t]
\centering
\vspace{-0.75\baselineskip}
\subfloat[]{
\label{fig:sim-ratio-gput}
\includegraphics[width=0.5\columnwidth]{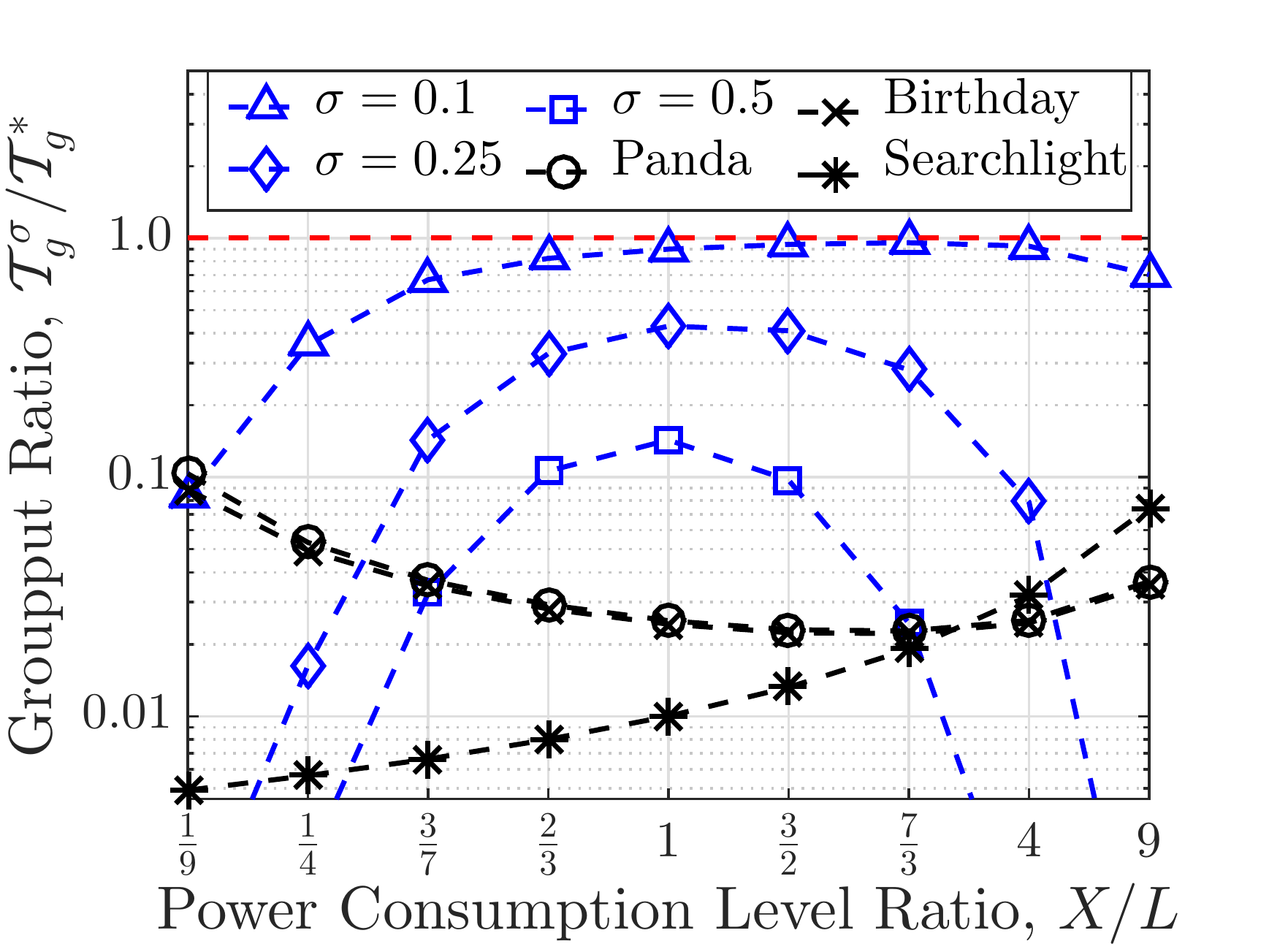}
}
\hspace{-18pt} \hfill
\subfloat[]{
\label{fig:sim-ratio-aput}
\includegraphics[width=0.5\columnwidth]{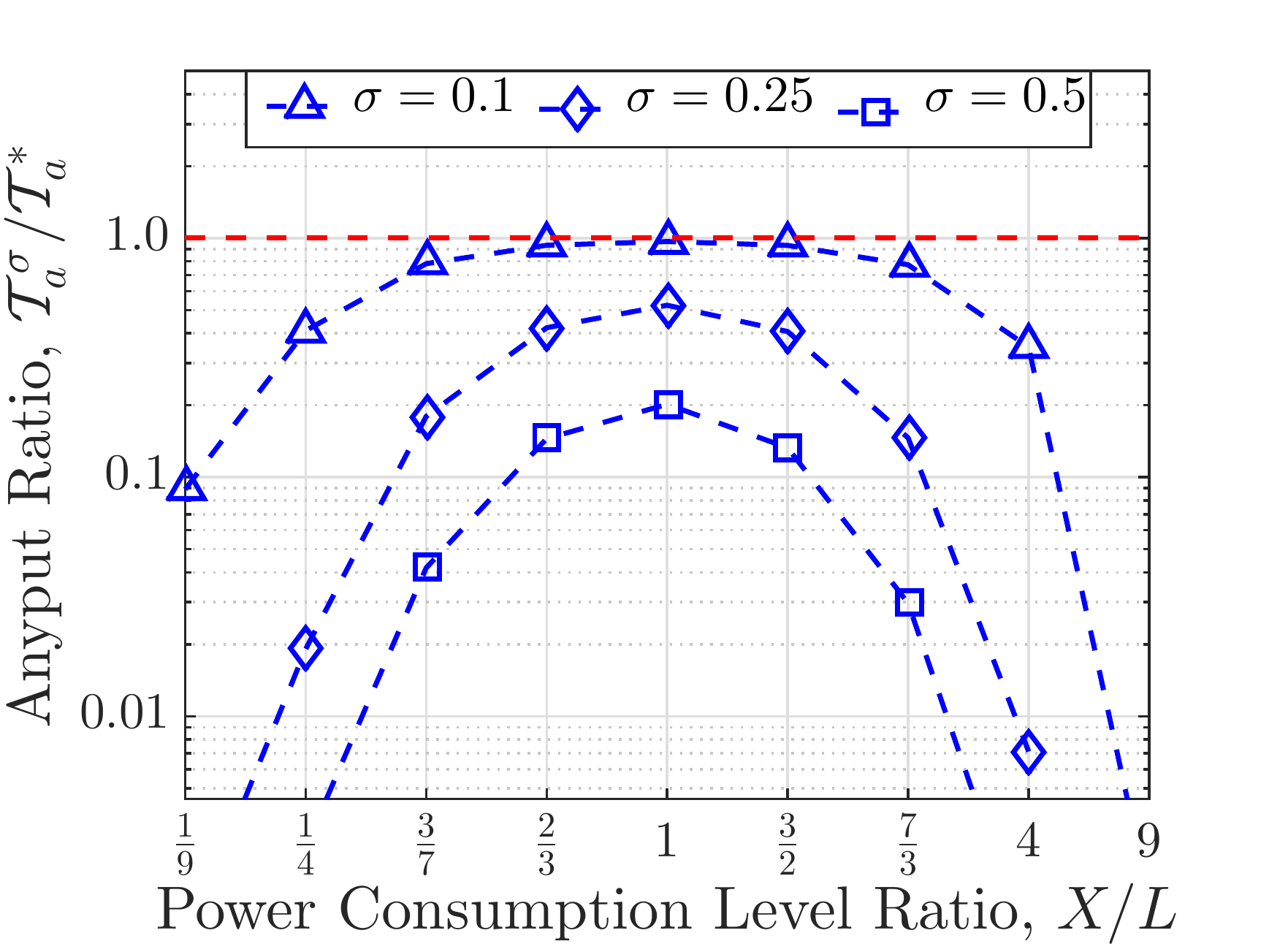}
}
\vspace{-0.5\baselineskip}
\caption{Throughput performance of {\namecaption} when operating in: (a) groupput mode and (b) anyput mode, with $\numnode = 5$, $\energybound{} = \ouruW{10}$, $\listencost{} + \xmitcost{} = \ourmW{1}$, and $\oursigma \in \{0.1, 0.25, 0.5\}$, as a function of $\xmitcost{} / \listencost{}$.}
\label{fig:sim-ratio-tput}
\vspace{-\baselineskip}
\end{figure}

Figs.~\ref{fig:sim-ratio-tput}\subref{fig:sim-ratio-gput} and~\ref{fig:sim-ratio-tput}\subref{fig:sim-ratio-aput} present, respectively, the groupput and anyput achieved by {\name} normalized to the oracle groupput and anyput, as a function of the power consumption ratio $\xmitcost{} / \listencost{}$, with $\numnode = 5$, $\energybound{} = \ouruW{10}$, and $\listencost{} + \xmitcost{} = \ourmW{1}$. Fig.~\ref{fig:sim-ratio-tput}\subref{fig:sim-ratio-gput} also presents the throughput achieved by Panda, Birthday, and Searchlight\footnote{For Searchlight protocol, we compare its throughput upper bound to $\UBG$.} protocols. The horizontal dashed lines at $1$ represent the oracle groupput and anyput.
Note that with $\listencost{} = \xmitcost{} = \ouruW{500}$, the ratio $\gputsigma / \UBG$ achieved by {\name} outperforms that of Panda by $6$x and $17$x with $\oursigma = 0.5$ and $\oursigma = 0.25$, respectively. In particular, the groupput ratio $\gputsigma / \UBG$ significantly outperforms that of prior art for $\xmitcost{} \approx \listencost{}$. The simulation results, which will be discussed later, also verify this throughput improvement.

Fig.~\ref{fig:sim-ratio-tput} shows that for a given value of $\xmitcost{} / \listencost{}$, $\tputsigma$ approaches $\UB$ with decreasing $\oursigma$, as expected (see Section~\ref{sec:distributed}).
Moreover, for each value of $\oursigma$, the throughput ratio $\tputsigma / \UB$ increases as the power consumption ratio $\xmitcost{} / \listencost{}$ is closer to $1$. This is realistic for current commercial low-power radios that have symmetric power consumption levels in listen and transmit states. This is due to the fact that (i) with small $\xmitcost{} / \listencost{}$ values, nodes enter transmit state infrequently, since listening is expensive and they must pass the listen state to enter the transmit state, and (ii) with large $\xmitcost{} / \listencost{}$ values, nodes spend energy transmitting even when there are no other nodes listening (e.g., $\numlisten(t)=0$). In particular, anyput degrades with large $\xmitcost{} / \listencost{}$ values, since anyput depends on the \textit{existence} of any active listeners when some node is transmitting. Therefore, when listening is expensive, the fact that multiple nodes listen simultaneously does not improve anyput. 
We believe that any distributed protocol will suffer from such performance degradation since, unlike Panda, Birthday, and Searchlight, nodes in a fully distributed setting \textit{do not} have any information about the properties of other nodes in the network.

\subsection{Burstiness and Latency}
\label{ssec:sim-burstiness}

The results until now suggest allowing $\oursigma \to 0$. While reducing $\oursigma$ improves throughput, it considerably increases the \emph{burstiness} of communication, as mentioned in Section~\ref{sec:distributed}. The burstiness is measured by the \emph{average burst length}, which is defined as the average number of packets that are successfully received in a burst (i.e., the average number of packets a node successfully receives before exiting listen state). The analytical computation of the average burst length can be found in Appendix~\ref{append:analysis-burstiness}.
In general, increased burstiness means that the long term throughput can be achieved with given power budgets but the variance of the throughput is more significant during short term intervals.
%
%
\begin{figure}[!t]
\centering
\vspace{-0.75\baselineskip}
\subfloat[]{
\label{fig:sim-burstiness-gput}
\includegraphics[width=0.5\columnwidth]{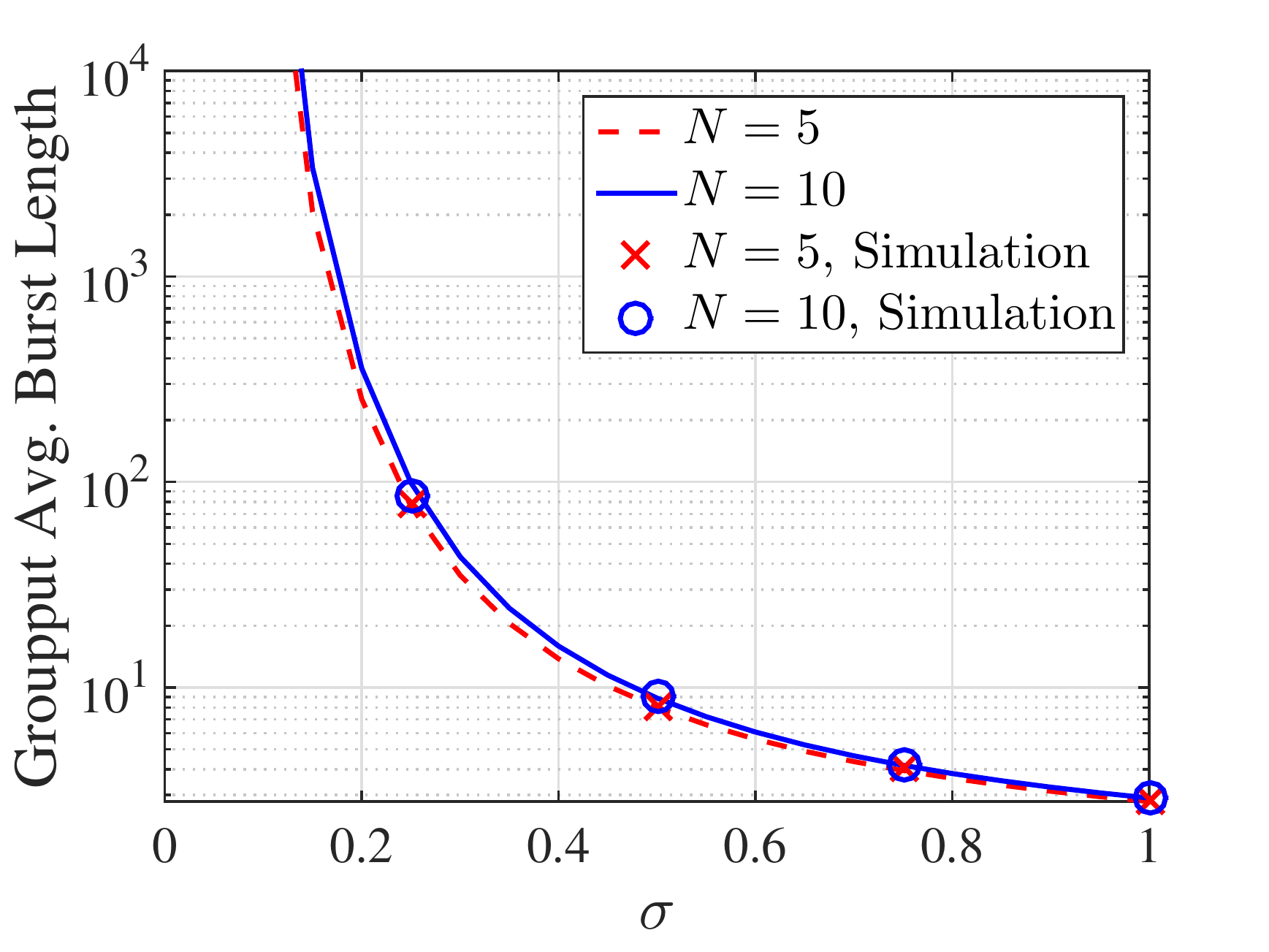}
}
\hspace{-18pt} \hfill
\subfloat[]{
\label{fig:sim-burstiness-aput}
\includegraphics[width=0.5\columnwidth]{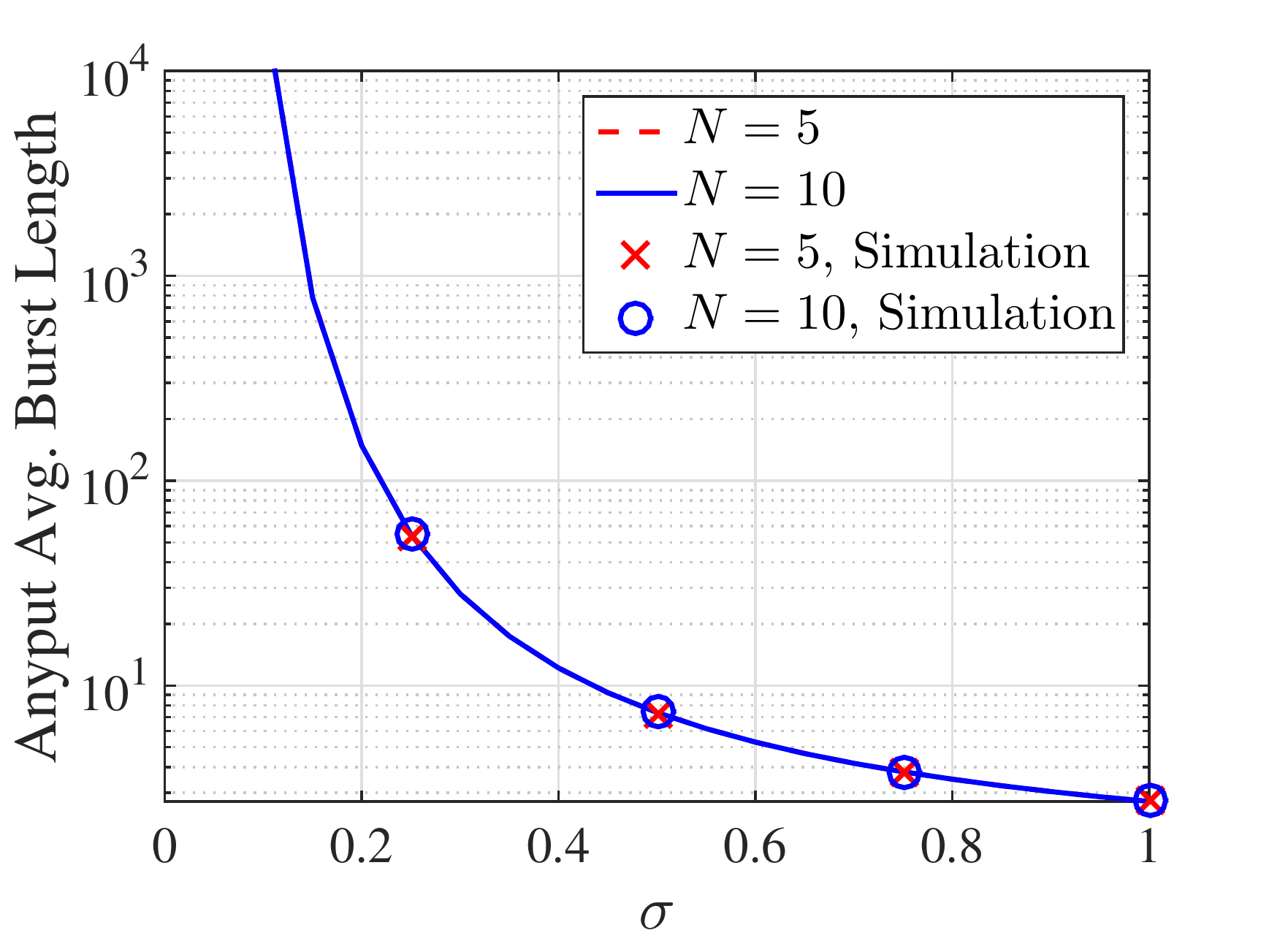}
}
\vspace{-0.5\baselineskip}
\caption{Analytical (curves) and simulated (markers) average burst length of {\namecaption} when operating in: (a) groupput mode and (b) anyput mode, with $\numnode \in \{5, 10\}$, $\oursigma \in \{0.25, 0.5\}$, $\energybound{} = \ouruW{10}$, and $\listencost{} = \xmitcost{} = \ouruW{500}$.}
\label{fig:sim-burstiness}
\vspace{-\baselineskip}
\end{figure}

Figs.~\ref{fig:sim-burstiness}\subref{fig:sim-burstiness-gput} and~\ref{fig:sim-burstiness}\subref{fig:sim-burstiness-aput} show the average burst length of {\name} (in log scale) when operating in groupput and anyput modes, respectively, in homogeneous networks with $\numnode \in \{5, 10\}$, $\energybound{} = \ouruW{10}$, $\listencost{} = \xmitcost{} = \ouruW{500}$, and varying $\oursigma$. Values are obtained using the analytical results~\eqref{eqn:analysis-burstiness-gput}--\eqref{eqn:analysis-burstiness-aput} derived in Appendix~\ref{append:analysis-burstiness} (curves) and contrasted with simulations at specific values of $\oursigma$ (markers).  Aside from showing that the simulation results and the analytical results are well matched, Fig.~\ref{fig:sim-burstiness} also demonstrates that reducing $\oursigma$ dramatically increases burstiness. For example, with $\oursigma = 0.25$ and $\numnode = 10$, a node running {\name} in groupput mode has an average burst length of $85$, and this value is increased to $4 \times 10^{5}$ with $\oursigma = 0.1$. This explains why $\gputsigmasim$ does not converge within reasonable time with $\oursigma = 0.1$ (see Section~\ref{ssec:sim-setup}). Comparing Fig.~\ref{fig:sim-burstiness}\subref{fig:sim-burstiness-gput} with Fig.~\ref{fig:sim-burstiness}\subref{fig:sim-burstiness-aput}, it can be seen that the groupput average burst length increases more rapidly than the anyput average burst length as $\oursigma$ decreases. Moreover, Fig.~\ref{fig:sim-burstiness}\subref{fig:sim-burstiness-aput} shows that the anyput average burst length is independent of $\numnode$, which corresponds to the analysis in Appendix~\ref{append:analysis-burstiness}. The reason is that the burst length of {\name} in anyput mode only depends on $\indisomelisten(t)$, which always equals to $1$, if the transmission is successful. We remark that reducing the burstiness is a subject of future work.


\begin{figure}[!t]
\centering
\vspace{-0.75\baselineskip}
\subfloat[]{
\label{fig:sim-latency-gput}
\includegraphics[width=0.5\columnwidth]{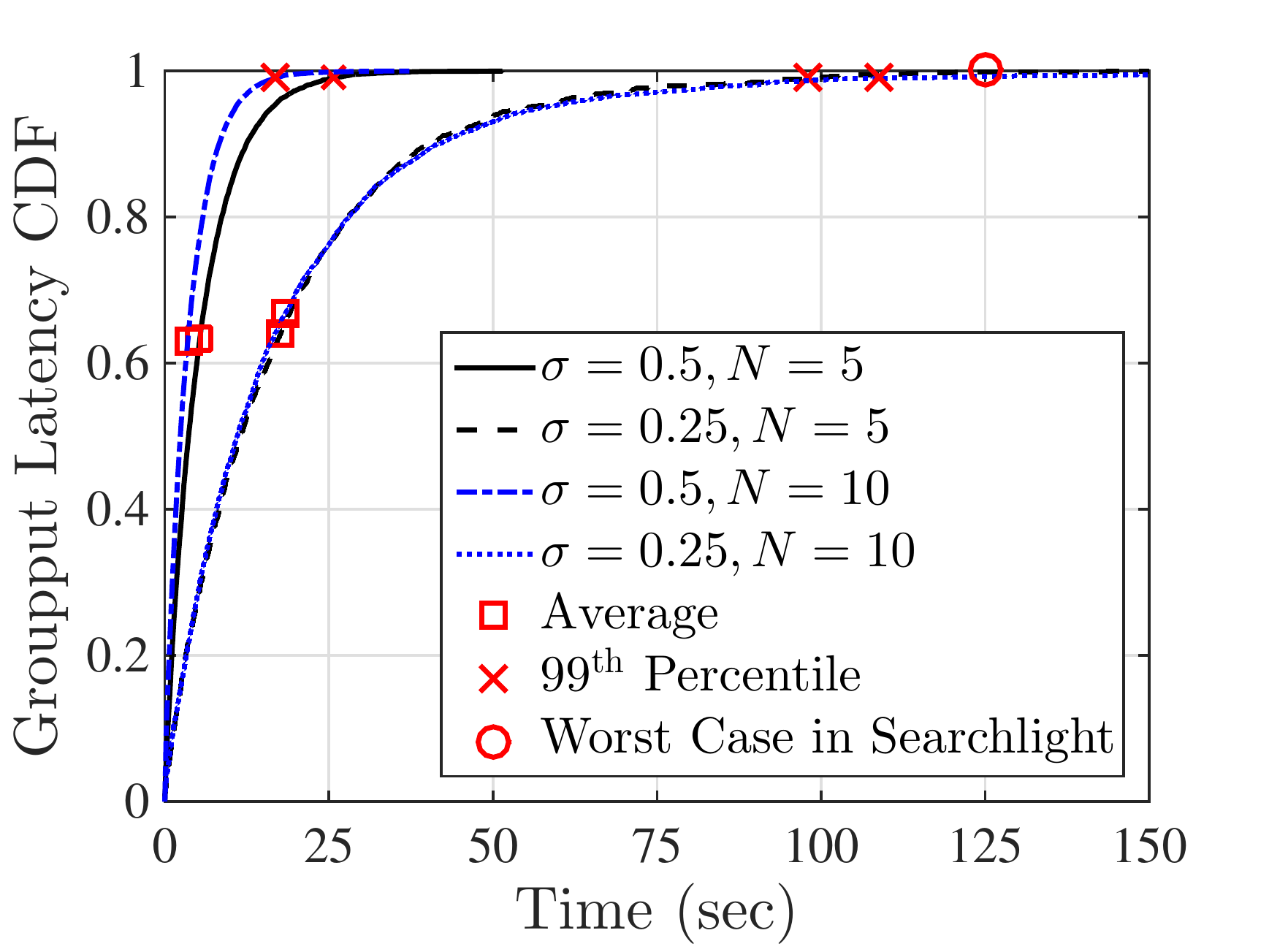}
}
\hspace{-18pt} \hfill
\subfloat[]{
\label{fig:sim-latency-aput}
\includegraphics[width=0.5\columnwidth]{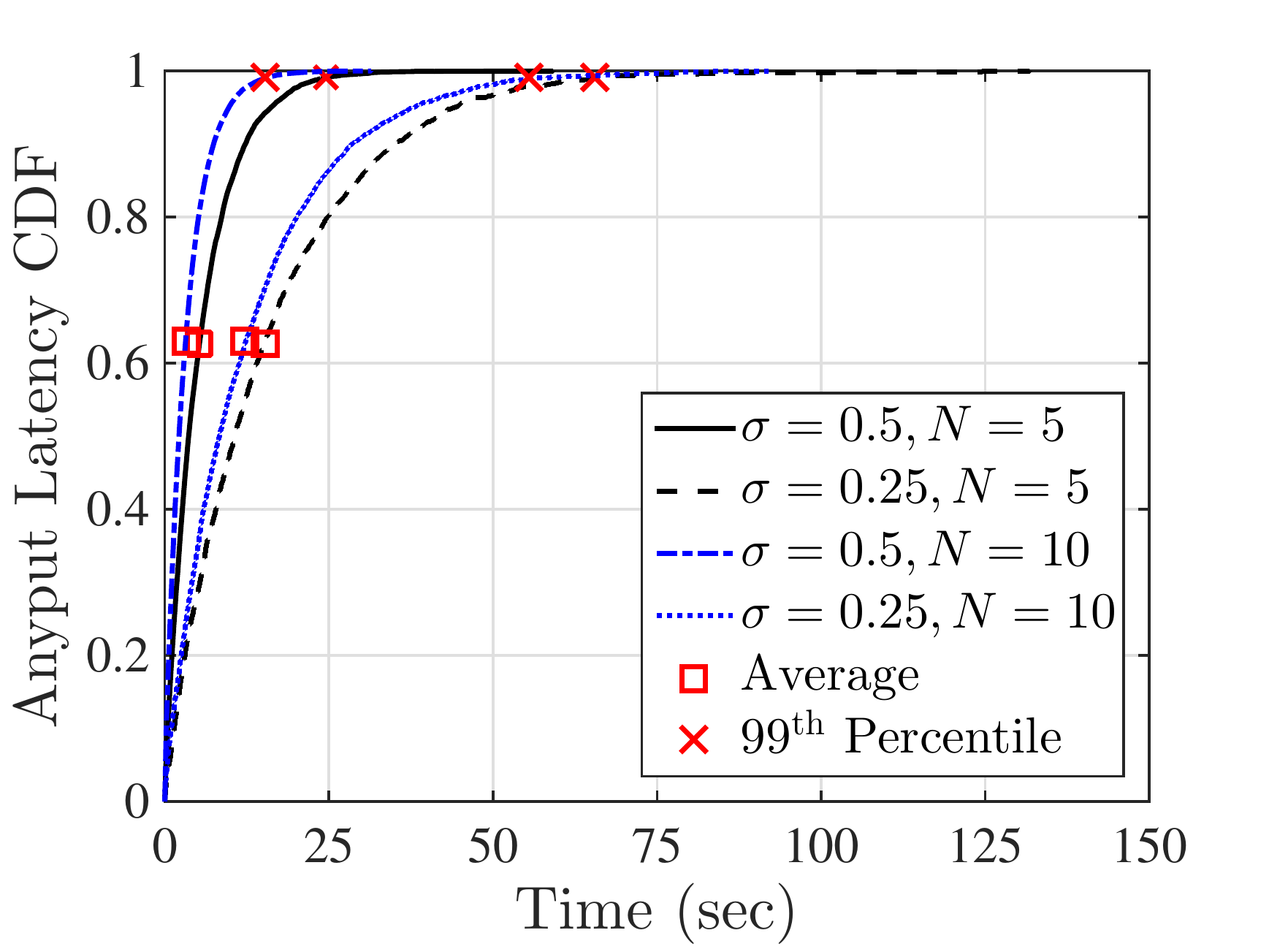}
}
\vspace{-0.5\baselineskip}
\caption{The CDF, mean, and $99^{\rm th}$ percentile latency of {\namecaption} when operating in: (a) groupput mode and (b) anyput mode, with $\numnode \in \{5, 10\}$, $\oursigma \in \{0.25, 0.5\}$, $\energybound{} = \ouruW{10}$, and $\listencost{} = \xmitcost{} = \ouruW{500}$.}
\vspace{-\baselineskip}
\label{fig:sim-latency}
\end{figure}

A second metric we consider is the communication \textit{latency}, which is defined as the time interval between consecutive bursts received by a node from some other node where the interval includes \textit{at least} one sleep period. We focus on this metric because nodes receiving longer bursts consume more energy, and therefore, need to sleep for longer periods of time.
Figs.~\ref{fig:sim-latency}\subref{fig:sim-latency-gput} and~\ref{fig:sim-latency}\subref{fig:sim-latency-aput} present the CDF latency of {\name} when operating in groupput and anyput modes obtained via simulations, and indicate both the average and the $99^{\rm th}$ percentile latency values. The homogeneous networks considered are with $\numnode \in \{5, 10\}$, $\oursigma \in \{0.25, 0.5\}$, $\energybound{} = \ouruW{10}$, and $\listencost{} = \xmitcost{} = \ouruW{500}$. Fig.~\ref{fig:sim-latency}\subref{fig:sim-latency-gput} also shows the pairwise worst case latency of Searchlight computed from~\cite{Bakht_mobicom2012} under the same power budget and consumption levels.\footnote{This is computed with slot length of $\ourms{50}$ and a beacon (packet) length of $\ourms{1}$ as was done in~\cite{sun2014hello}.}

Fig.~\ref{fig:sim-latency} shows that the latency increases as $\oursigma$ decreases, since nodes receiving more packets in a short time period (i.e., increased burstiness) have higher variation in their energy storage levels, and need to sleep longer to restore energy.
Fig.~\ref{fig:sim-latency} also shows that a larger value of $\numnode$ results in lower latency, since every node is more likely to receive when more nodes exist in the network.
Comparing Fig.~\ref{fig:sim-latency}\subref{fig:sim-latency-gput} with Fig.~\ref{fig:sim-latency}\subref{fig:sim-latency-aput}, it is observed that {\name} operating in anyput mode has slightly lower average latency than in groupput mode. However, with a smaller $\oursigma$ value (i.e., $\oursigma = 0.25$), the $99^{\rm th}$ percentile latency of {\name} when operating in anyput mode is significantly lower than that in groupput mode. This results from the fact that the average burst length of {\name} in anyput mode depends on the existence of any listening nodes, whose value is always less than or equal to the number of listening nodes considered in groupput mode.

For all parameters considered, the $99^{\rm th}$ percentile groupput latency is within $120$ seconds, outperforming the Searchlight pairwise worst case latency bound of $125$ seconds. Note that although {\name} has a non-zero probability of having any latency, its latency is below the worst case latency of Searchlight in most cases (over $99\%$).


\subsection{Groupput Evaluation in Non-clique Topologies}
\label{ssec:sim-nonclique}

\begin{figure}[!t]
\begin{center}
\includegraphics[width=0.5\columnwidth]{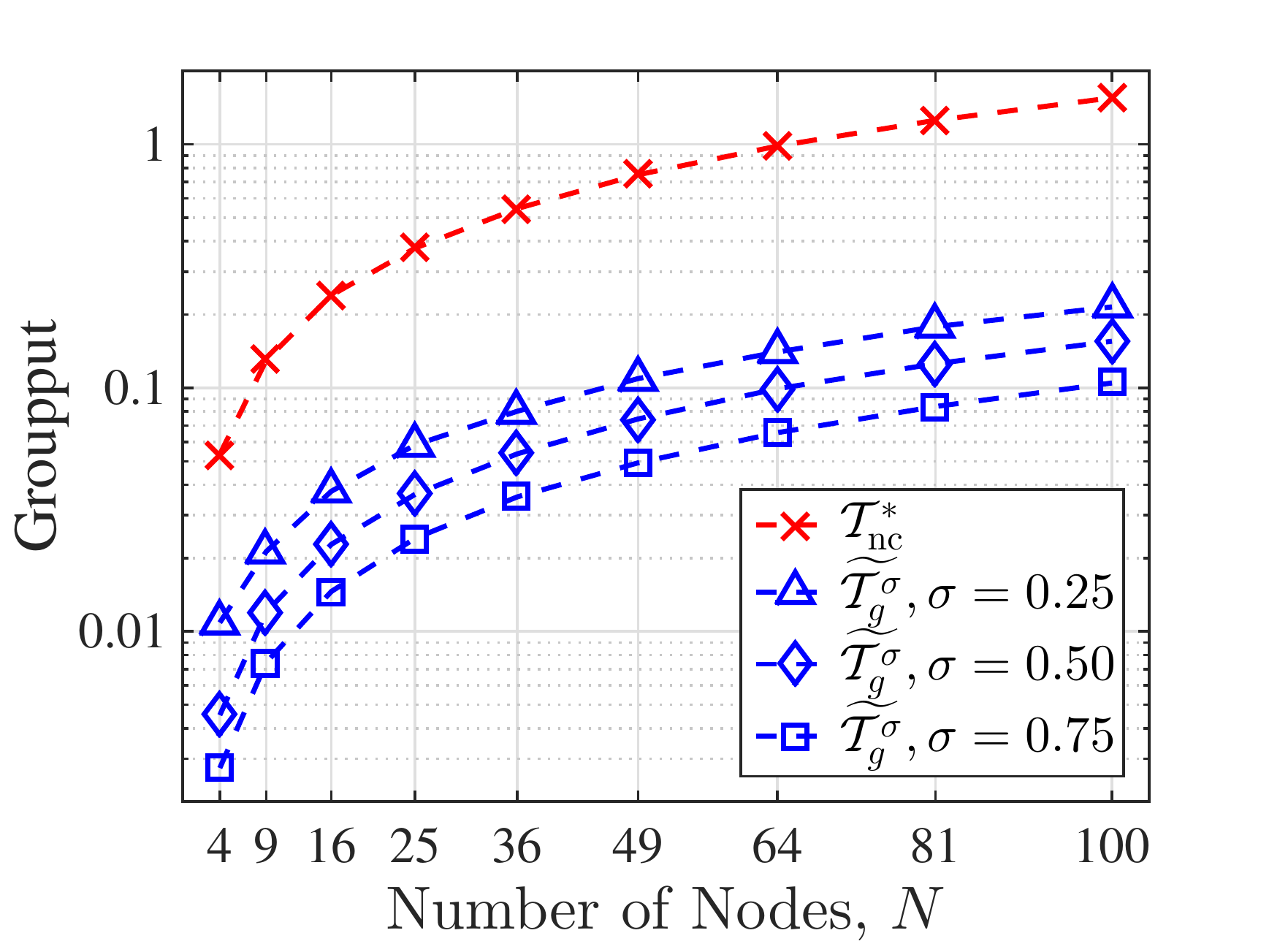}
\caption{The oracle groupput, $\UBNC$, and the throughput of {\namecaption} when operating in groupput mode obtained via simulations, $\gputsigmasim$, in grid topologies with varying $\numnode$, $\oursigma \in \{0.25, 0.5, 0.75\}$, $\energybound{} = \ouruW{10}$, and $\listencost{} = \xmitcost{} = \ouruW{500}$.}
\label{fig:sim-nonclique-gput}
\vspace{-1\baselineskip}
\end{center}
\end{figure}

We now compute the oracle groupput for non-clique topologies (derived in Section~\ref{ssec:upper-additional}) and evaluate the groupput achieved by {\name} in such scenarios via simulations.
Since simultaneous transmissions can happen in non-clique topologies, none of the transmissions will be counted as throughput in the simulations. 

We use grid topologies with varying number of nodes, $\numnode$, in which each node has at most $4$ neighbors. For example, $\numnode = 25$ represents a $5 \times 5$ grid. Fig.~\ref{fig:sim-nonclique-gput} presents the oracle groupput, $\UBNC$, for grid topologies, and the throughput achieved by {\name} in groupput mode obtained via simulations with with varying $\oursigma$ and $\numnode$. Note that for all the grid topologies considered, the upper and lower bounds of $\UBNC$ (see Section~\ref{ssec:upper-additional}) are the same, providing the exact oracle groupput.

Fig.~\ref{fig:sim-nonclique-gput} shows that {\name} achieves $14\% - 22\%$ of the oracle groupput, $\UBNC$, with $\oursigma = 0.25$. Although increasing $\oursigma$ leads to lower groupput, it can be observed that as $\numnode$ increases, the groupput approaches $10\%$ of $\UBNC$ with $\oursigma = 0.5$. Despite the fact that the groupput cannot be obtained for $\oursigma = 0.1$, achieving $10\% - 20\%$ of $\UBNC$ is remarkable given the fact that {\name} operates in a distributed manner.

\section{Experimental Evaluation}
\label{sec:impl}

To experimentally evaluate the performance of {\namecap},\footnote{See Section~\ref{ssec:distributed-variants} the reasons for only implementing {\namecapcaption}.} we implement it using the Texas Instruments eZ430-RF2500-SEH node~\cite{instruments2013ez430}.\footnote{A demonstration of the testbed is presented in~\cite{chen_sensys2016demo}.} In this section, we first describe the energy measurements performed on the nodes running {\namecap}. Then, we describe the method by which nodes can estimate the number of listening nodes. Finally, we experimentally evaluate the performance of {\namecap}.

\subsection{Experimental Setup}
The TI eZ430-RF2500-SEH node is equipped with: (i) an ultra-low-power MSP430 microcontroller and a CC2500 wireless transceiver operating at $\SI{2.4}{\giga\hertz}$ at $\SI{250}{\Kbps}$, (ii) a solar energy harvester (SEH-01) that converts ambient light into electrical energy, and (iii) a $\ourmF{1}$ capacitor to power up the transceiver board. Despite its drawbacks which will be discussed below, it can be used for evaluation by extending the length of the shortest allowable data transmission.

We consider power budgets of $\energybound{} \in \{ \ourmW{1}, \ourmW{5} \}$. From our measurements, a node spends $\listencost{} = \ourmW{67.08}$ in the listen state and $\xmitcost{} = \ourmW{56.29}$ in the transmit state.\footnote{This corresponds to a $\SI{-16}{\dBm}$ transmission power, at which nodes within the same room typically have little or no packet loss.} The power consumption levels are very similar from node to node. Recall from Section~\ref{sec:sim} that the performance of {\name} depends on the ratio between the power consumption levels and budget. Therefore, our experimental results will be similar to experiments when both the power consumption levels and budget are scaled down (e.g., a network of nodes with $\energybound{} \in \{\ouruW{10}, \ouruW{50} \}$, $\listencost{} = \ouruW{670}$, and $\xmitcost{} = \ouruW{560}$).

Each node is programmed with its $\energybound{}$, $\listencost{}$, and $\xmitcost{}$ as the input of {\namecap}.
The nodes' main drawbacks include (i) inaccurate readings of the energy storage level (i.e., the voltage of the on-board capacitor) which are sensitive to the environment, and (ii) the fact that the $\ourmF{1}$ capacitor cannot support multiple packet transmissions.  Due to these drawbacks, we implement (via software) a virtual battery at each node.   The virtual battery emulates the node's energy storage level based on its sleep, listen, and transmit activities, and is used for updating the Lagrange multiplier according to (\ref{eqn:update-lm-epoch}). We show in the following section that in practice, a node running {\namecap} using this virtual battery is indeed consuming power at a rate close to its power budget.

\subsection{Energy Consumption Measurements}
\label{ssec:impl-energy-meas}

To accurately measure the power consumption of the nodes, we disable the on-board solar cell, and attach a large pre-charged capacitor ($\capsize = \ourF{5}$) that stores energy in advance. 
The energy consumed is computed by
\begin{equation}
\label{eqn:impl-compute-total-energy-consumption}
E_{\rm consumed} = 0.5 \capsize \cdot \left( V_{t_0}^2 - V_{t_1}^2 \right),
\end{equation}
where $V_{t_0}$ and $V_{t_1}$ are the measured power voltage values of the capacitor at $t_0$ and $t_1$. The empirical average power consumption, $\avgconsumerateexp (\ourmW{})$, is then computed by
\begin{equation}
\label{eqn:impl-compute-avg-energy-consumption}
\avgconsumerateexp = E_{\rm consumed} / \left( t_1 - t_0 \right).
\end{equation}
Note that even with such a big capacitor, a node with a power budget of $\ourmW{1}$ ($\ourmW{5}$) has a lifetime of only $135$ ($27$) minutes with $V_{t_0} = \ourV{3.6}$ and $V_{t_1} = \ourV{3.0}$, which represent its stable working voltage range. 

To measure the power consumption of the nodes, we charge the capacitor to $V_{t_0} = \ourV{3.6}$ and log the readings of $V_{t_1}$ after $30$ minutes using a multimeter. The empirical average power consumption is computed from (\ref{eqn:impl-compute-total-energy-consumption}) and (\ref{eqn:impl-compute-avg-energy-consumption}) for $\oursigma \in \{0.25, 0.5\}$ and is averaged using $60$ runs. Because $\listencost{}$ and $\xmitcost{}$ do not account for some additional energy usage,\footnote{The additional energy usage includes the energy consumed in powering up the regulator circuitry, etc.} the {\em actual power consumption}, $\avgconsumerateexp$, is in fact a small fraction higher than the {\em target power budget}, $\energybound{}$. Irrespective of $\oursigma$, the measurement results show that $\avgconsumerateexp$ exceeds $\energybound{}$ by $11\%$ for $\energybound{} = \ourmW{1}$, and by $4\%$ for $\energybound{} = \ourmW{5}$.

Observing the empirical power consumption of the nodes, we compute the achievable throughput by solving {\namecp} using both the actual power consumption, $\avgconsumerateexp$, and the target power budget, $\energybound{}$, denoted by $\overline{\gputsigma}$ and $\gputsigma$, respectively. In Section~\ref{ssec:impl-eval}, we compare the experimental throughput to both $\overline{\gputsigma}$ and $\gputsigma$.
Having verified the power consumption of the nodes, we replace the capacitor with AAA batteries,\footnote{The constant voltage of AAA batteries limits the ability to measure the power consumption of the nodes.} allowing the experiments to run for longer times.

\begin{figure*}[!htb]
\centering
\hspace{-0.10in} 
\includegraphics[width=0.54\columnwidth]{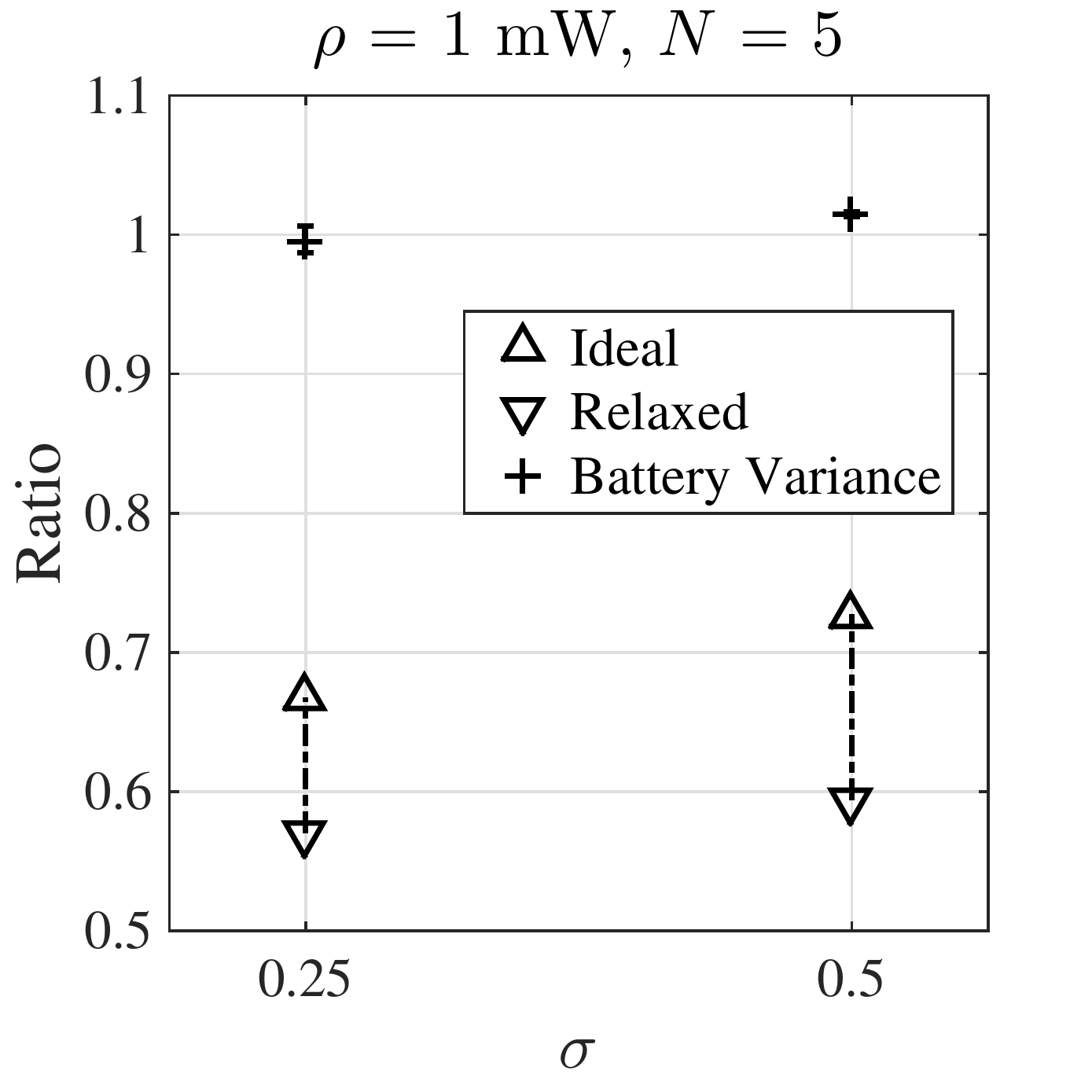} \hspace{-18pt} \hfill \includegraphics[width=0.54\columnwidth]{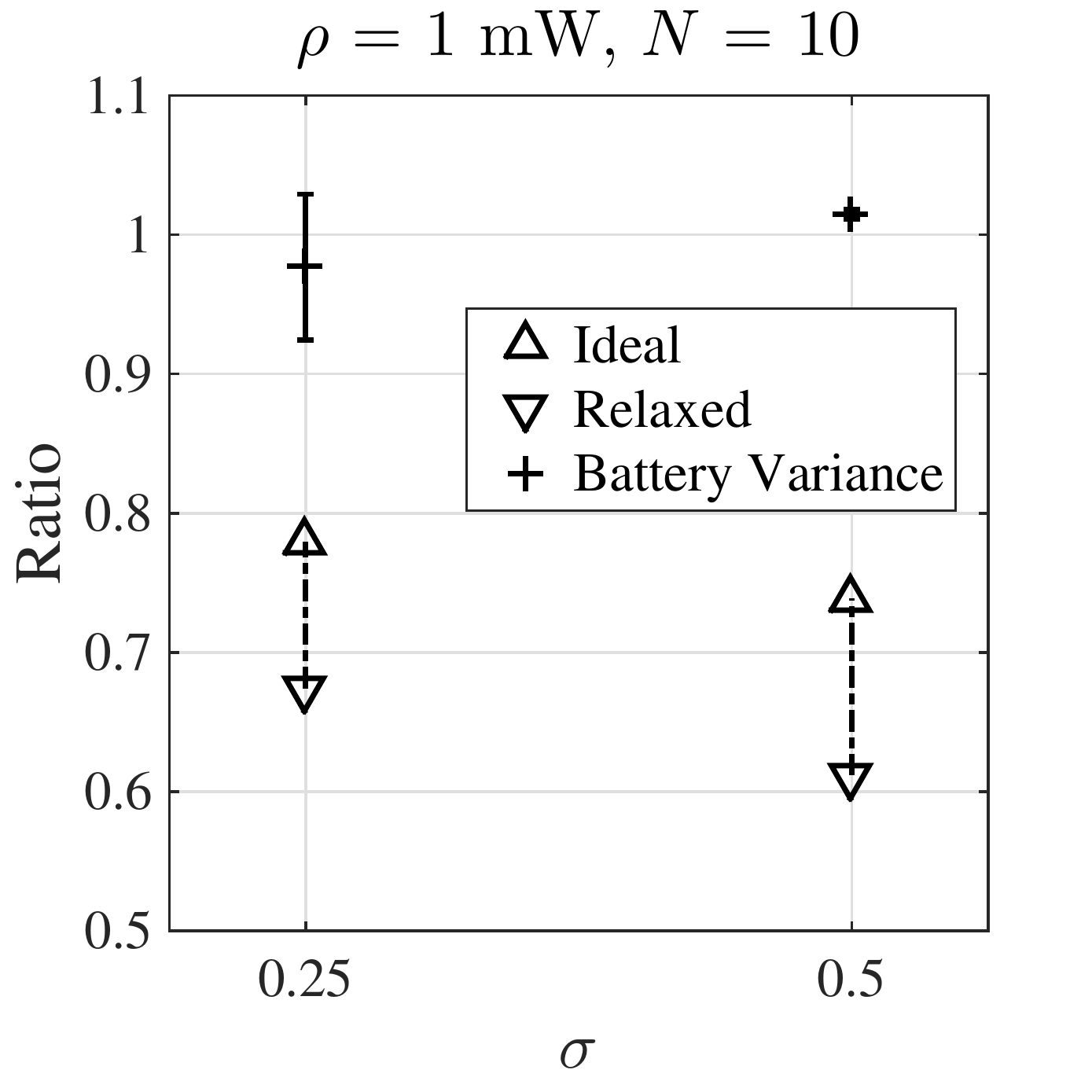} \hspace{-18pt} \hfill
\includegraphics[width=0.54\columnwidth]{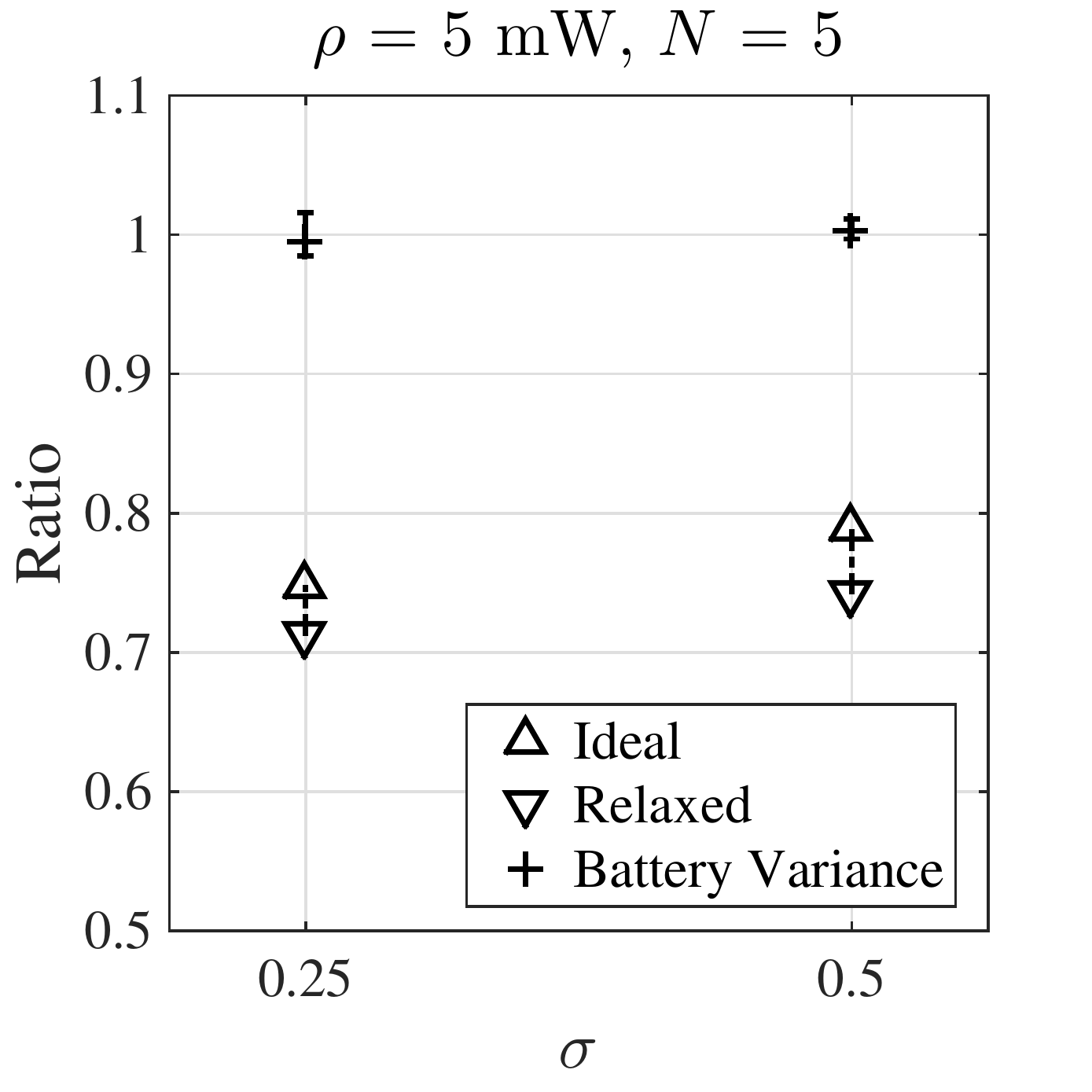} \hspace{-18pt} \hfill \includegraphics[width=0.54\columnwidth]{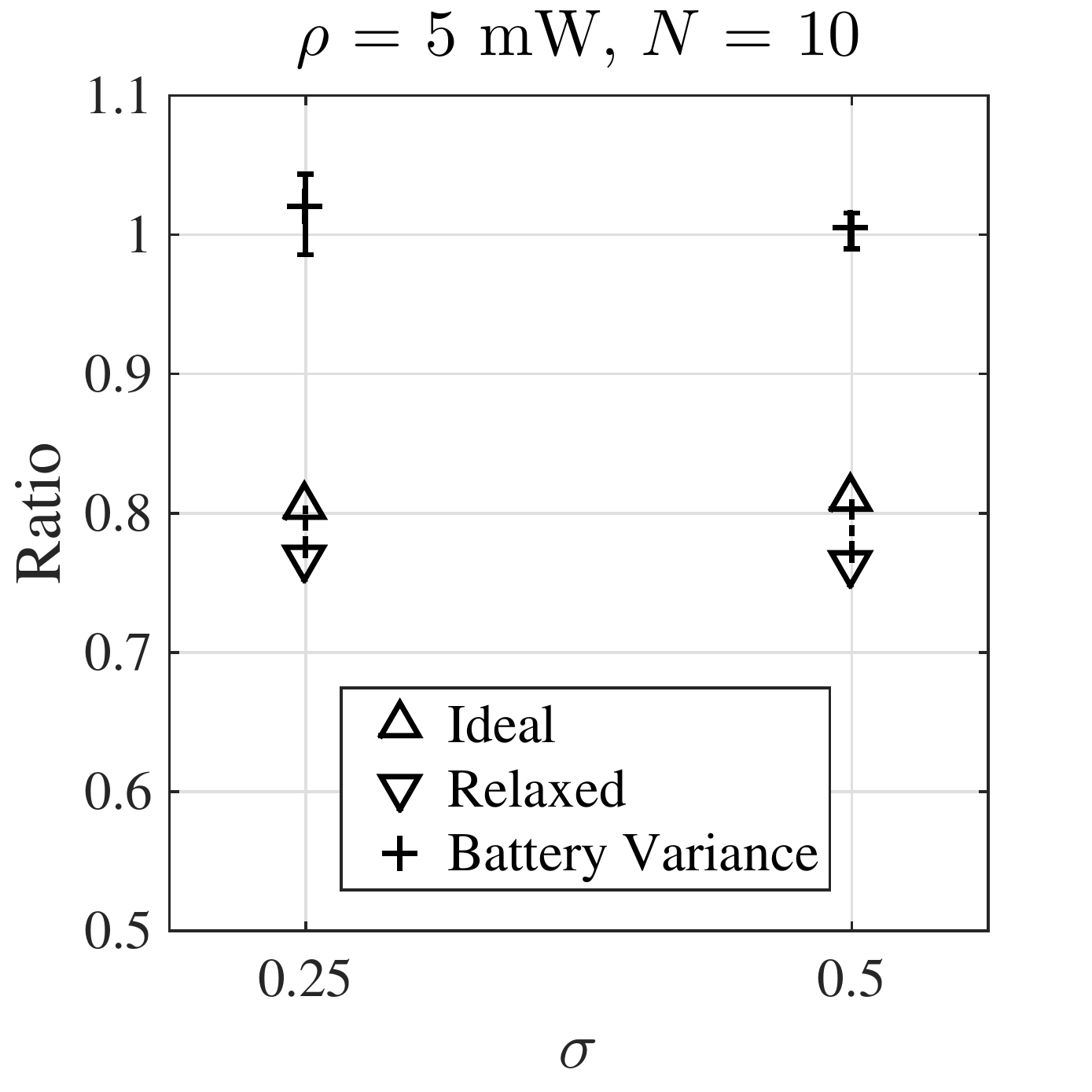}
\caption{Points marked as ``Ideal'' (``Relaxed'') represent ratio of experimental throughput normalized to the achievable throughput obtained by using the target power budget (actual power consumption) and points marked as ``Battery Variance'' present the average, minimum, and maximum ratios of power consumption normalized to target power budget, with $\numnode \in \{5, 10\}$, $\energybound{} \in \{ \ourmW{1}, \ourmW{5} \}$, and $\oursigma \in \{0.25, 0.5\}$.}
\label{fig:impl-tput}
\vspace{-0.5\baselineskip}
\end{figure*}

\subsection{Practical Pinging}
\label{ssec:impl-practical-ping}

To enable practical pinging in {\namecap}, a short, fixed-length {\em pinging interval} is introduced after each packet transmission. During this interval, the transmitter listens for pings and recipients of the previous packet send a short ping at a random time uniformly distributed within the interval. The transmitter then estimates the number of listeners, $\numlistenest(t)$, by counting the pings it receives, and adjusts the transition rate, $\transXL^C(t)$, according to~\eqref{eqn:trans-rate-xl-cap}.

Ideally, each ping should be much shorter than both the pinging interval and the packet length in order to reduce the collisions between pings, as well as for the transmitter to successfully receive it. Therefore, we use pings of length $\ourms{0.4}$, which is the shortest packet that can be sent by a node. Based on this, we empirically set the pinging interval to $\ourms{8}$ and each data packet to $\ourms{40}$.


\subsection{Performance Evaluation}
\label{ssec:impl-eval}

We consider homogeneous networks with $\numnode \in \{5, 10\}$, $\energybound{} \in \{\ourmW{1}, \ourmW{5}\}$,  $\oursigma \in \{0.25, 0.5\}$, and nodes are located in proximity. One additional listening node (a $6^{\rm th}$ or $11^{\rm th}$ node) is also present but only as an observer and is connected to a PC via a USB port. Each data packet contains the node ID and information about the number of packets it has received from each other node.
The observer node reports all received packets to the PC for storage and post processing.
Each experiment is conducted for up to $24$ hours. The experimental throughput is computed by dividing the duration of successful transmissions by the experiment duration.

\noindent{\bf Throughput evaluation}: Fig.~\ref{fig:impl-tput} presents the ratio of the experimentally obtained throughput, $\gputsigmaexp$, normalized to the achievable throughput $\gputsigma$ and $\overline{\gputsigma}$ (see Section~\ref{ssec:impl-energy-meas}). Separate charts represent the results for differing number of nodes, $\numnode$, and power budget, $\energybound{}$.
Points marked ``Ideal'' show the experimental throughput normalized to the achievable throughput computed by solving {\namecp} with the target power budget $\energybound{}$ (i.e., $\gputsigmaexp / \gputsigma$). Points marked ``Relaxed'' show the experimental throughput normalized to the achievable throughput computed by solving {\namecp} with the actual power consumption $\avgconsumerateexp$ (i.e., $\gputsigmaexp / \overline{\gputsigma}$). As expected, $\overline{\gputsigma}$ is higher than $\gputsigma$, resulting in a lower throughput ratio.

Fig.~\ref{fig:impl-tput} shows that despite the practical limitations (e.g., packet collisions and inaccurate clocks) faced when running {\namecap} on real hardware, the ratio $\gputsigmaexp / \overline{\gputsigma}$ is between $57\% - 77\%$ ($\gputsigmaexp / \gputsigma$ is between $67\% - 81\%$) for all settings considered.
Moreover, Table~\ref{table:impl-tput-comparison} shows the improvement of {\namecap} over the throughput of Panda computed according to~\cite{PANDA}, denoted by $\tput_{\rm Panda}$, under the same power consumption levels and budget, with $\oursigma = 0.25$. It can be seen that with power budget of $\energybound{} = \ourmW{1}$, the experimental throughput of {\namecap} outperforms the analytically computed throughput of Panda by $8$x -- $11$x.

We remark that getting a higher experimental throughput ratio is limited by the following reasons. First, there is an $\ourms{8}$ pinging interval (see Section~\ref{ssec:impl-practical-ping}) after each packet transmission which effectively reduces the number of bits delivered. Second, collisions of pings or failed decodings of pings result in inaccurate estimates of the number of listeners. Third, the low-power clock used by a node during its sleep state drifts and additionally can be affected by its environment.

\begin{table}[!t]
\caption{Experimental throughput of {\namecapcaption} compared to computed throughput of Panda (all normalized to the achievable throughput $\gputsigma$), with $\oursigma = 0.25$ and varying $(\numnode, \energybound{})$.}
\label{table:impl-tput-comparison}
\vspace{-0.5\baselineskip}
\small
\begin{center}
\begin{tabular}{|c|c|c|c|c|}
\hline
$(\numnode, \energybound{}(\ourmW{}))$ & $(5,1)$ & $(10,1)$ & $(5,5)$ & $(10,5)$ \\
\hline
$\gputsigmaexp / \gputsigma ~(\%)$ & $66.78$ & $77.96$ & $74.84$ & $80.53$ \\
\hline
$\tput_{\rm Panda} / \gputsigma ~(\%)$ & $6.24$ & $9.64$ & $19.35$ & $35.63$ \\
\hline
$\gputsigmaexp / \tput_{\rm Panda}$ & $10.76$ & $8.09$ & $3.87$ & $2.26$ \\
\hline
\end{tabular}
\vspace{-1\baselineskip}
\end{center}
\normalsize
\end{table}
\noindent{\bf Power consumption}:
In Section~\ref{ssec:impl-energy-meas}, we show that the power consumption of the virtual battery is valid for evaluating the actual power consumption of the node. Fig.~\ref{fig:impl-tput} also presents the mean, minimum, and maximum power consumption of the virtual battery normalized to the target power budget $\energybound{}$. Specifically, a value of $1$ means that a node consumes power on average at the rate of its power budget throughout the experiment, and a higher value means that a node consumes power at the rate which is higher than its power budget.

The results show that nodes running {\namecap} consume power at rates which are within $7\%$ and $3\%$ of the target power budget with $\oursigma = 0.25$ and $\oursigma = 0.5$, respectively. This is because smaller value of $\oursigma$ increases the communication burstiness (see Section~\ref{ssec:sim-burstiness}), resulting in larger variance of the nodes' virtual battery levels.

\noindent {\bf Collection of Pings}: An important input to {\namecap} is the estimates of number of active listeners, $\numlistenest(t)$, based on which the transmitter decides the probability to continuously transmit. Larger values of $\numlistenest(t)$ lead to longer average burst length and can potentially significantly increase the throughput. For example, receiving $1$ ping, the transmitter continuously transmits a packet with probability $0.8647$ with $\oursigma = 0.5$. This probability increases to $0.9817$ with $\oursigma = 0.25$, which substantially increases the burstiness. Also, with lower power budget, a successful transmission happens more rarely and it becomes harder to collect pings.

Table~\ref{table:impl-avg-pings} presents the distribution of number of pings (i.e., number of active listeners) received by the transmitter after each packet transmission, during experiments of $\numnode = 5$, $\oursigma = 0.25$, and $\energybound{} \in \{\ourmW{1}, \ourmW{5}\}$. It can be shown that with a higher power budget, the nodes are more active and the transmitter has higher probability to receive more pings. On the other hand, with lower power budget, the transmitter almost never receives more than $3$ pings in a $5$ nodes experiment, resulting in lower throughput as illustrated in Fig.~\ref{fig:impl-tput}.

\begin{table}[!t]
\caption{Distribution of number of pings (active listeners) received after each packet transmission with $\numnode = 5$, $\oursigma = 0.25$, and varying $\energybound{}$.}
\label{table:impl-avg-pings}
\vspace{-0.5\baselineskip}
\small
\begin{center}
\begin{tabular}{|c|c|c|c|c|c|}
\hline
\# of Listeners & $0$ & $1$ & $2$ & $3$ & $4$ \\
\hline
$\energybound{}=\ourmW{1}(\%)$ & $89.03$ & $9.69$  & $1.28$ & $0.00$ & $0.00$ \\
\hline
$\energybound{}=\ourmW{5}(\%)$ & $59.21$ & $31.22$ & $8.22$ & $1.24$ & $0.11$ \\
\hline
\end{tabular}
\vspace{-1\baselineskip}
\end{center}
\normalsize
\end{table}


\section{Conclusion}
\label{sec:conclusion}
In this paper, we considered the problem of maximizing the broadcast groupput and anyput among a set of energy-constrained nodes with heterogeneous power budgets and listen and transmit power consumption levels. We also provided efficient methods to obtain oracle groupput and oracle anyput for a given set of heterogeneous nodes.

We developed the {\name} distributed protocols that control the nodes' transitions among sleep, listen, and transmit states. We analytically showed that heterogeneous nodes using the protocols (without any a priori knowledge regarding the number of nodes in the network, and power budgets and consumption levels) can achieve the oracle groupput and anyput in a limiting sense (when $\oursigma \to 0$).

We evaluated the throughput performance of {\name} numerically and through extensive simulations, and compared it to the state of the art. We also considered the design tradeoffs in relation to $\oursigma$ and the impact of $\oursigma$ on the burstiness and throughput. Finally, we experimentally evaluated {\name} using commercial-off-the-shelf energy harvesting nodes, thereby demonstrating its practicality.

There are several open future research directions. In particular, future research will focus on extending the analysis to non-clique toplogies. Moreover, evaluation with custom-designed ultra-low-power nodes (e.g.,~\cite{Margolies_EnHANTS_TOSN}), that have improved energy awareness compared to the TI eZ430-RF2500-SEH nodes, would enable to better assess the tradeoffs related to the protocol design. Finally, considering unique application characteristics and their relation to groupput and anyput is an open problem. 



\bibliographystyle{IEEEtran} 
\bibliography{econcast}

\appendices
\section{Proof of Lemma~\ref{lem:lp-schedule}}
\label{append:proof-lp-schedule}

We describe here a schedule that works assuming each packet transmission has a fixed transmit length of $\slotlength$, though the proof can be extended to varying transmission lengths. Therefore, time can be broken into slots of length $\theta$ (by the oracle) and nodes sleep, listen, or transmit on a per-slot basis.

Assume that the optimal solution $\optsolvec$ to {\namelpgput} yields rational values for all $\listenfrac{i}^*$ and $\xmitfrac{i}^*$. 
The period size of the oracle schedule $\periodsize$ is set to the least common denominator over all solution variables in $\optsolvec$. Hence, during each period, node $i$ listens for $\listenfrac{i}^* \periodsize$ slots and transmits for $\xmitfrac{i}^* \periodsize$ slots. The slots during the period can be assigned arbitrarily by the oracle to transmitters (e.g., weighted round-robin or in-order) and~\eqref{eqn:non-overlap-transmit} ensures that there will be sufficient slots. Once the slots for the transmitters are assigned, each listener can then choose their $\listenfrac{i}^{*} \periodsize$ slots in which they listen from the set of transmit slots assigned to other transmitters, and~\eqref{eqn:listen-when-xmit} ensures these are sufficient as well (note that multiple listeners are permitted for a single transmitter slot).

If the periodic scheduler is launched immediately, some nodes may not have the harvested (or budgeted) energy to perform all listen and transmit tasks within the first period (e.g., it may be assigned to transmit and listen early on, and recoup the energy during later slots). If such a case, we simply delay the initial iteration, allowing all nodes to harvest (or budget) $\energybound{i} \periodsize \slotlength$ amount of energy for that initial period. Then the nodes have enough energy to repeat all the subsequent periods since the energy node $i$ spends during the $k$-th period is $(\listenfrac{i} \listencost{i} \periodsize \slotlength + \xmitfrac{i} \xmitcost{i} \periodsize \slotlength)$, and the energy it accumulates during this period is $\energybound{i} \periodsize \slotlength$. Hence, it follows from~\eqref{eqn:energy-constraint} that no more energy is spent than is accumulated (budgeted). Therefore, there is sufficient energy to repeat the period. \hfill $\blacksquare$

\section{Solving {\namelpgput} and {\namelpaput}}
\label{append:proof-lp-homo-soln}

The optimization problem {\namelpgput} can be solved via two steps. First, we show that in the optimal solution to {\namelpgput}, the equalities strictly hold in constraints~\eqref{eqn:energy-constraint} and~\eqref{eqn:listen-when-xmit}. Next, we solve for the optimal solution. Note that in homogeneous networks, the constraints become $\listenfrac{} \listencost{} + \xmitfrac{} \xmitcost{} \leq \energybound{}$ and $\listenfrac{} \leq (\numnode-1) \xmitfrac{}$, and $\listenfrac{i}^* = \listenfrac{}^{*}$ and $\xmitfrac{i}^{*} = \xmitfrac{}$.

We prove the first part by contradiction. Note that if both inequalities strictly hold, $\listenfrac{}$ can always be increased, resulting in higher throughput. Therefore, if suffices that at least one of the two inequalities is satisfied with strict equality.

\noindent{\bf Case 1}: If $\listenfrac{}^{*} \listencost{} + \xmitfrac{}^{*} \xmitcost{} = \energybound{} - \epsilon$ for some $\epsilon > 0$ and $\listenfrac{}^{*} = (\numnode-1) \xmitfrac{}^{*}$, the optimal solution is given by $\xmitfrac{}^{*} = \frac{\energybound{} - \epsilon}{\xmitcost{} + (\numnode-1) \listencost{}}, ~\listenfrac{}^{*} = (\numnode-1) \xmitfrac{}^{*}$. Let a new solution be
\begin{align*}
\listenfrac{}' = \listenfrac{}^{*} + \frac{(\numnode-1) \epsilon}{\xmitcost{} + (\numnode-1) \listencost{}}, ~\xmitfrac{} ' = \xmitfrac{}^{*} + \frac{\epsilon}{\xmitcost{} + (\numnode-1) \listencost{}},
\end{align*}
and it can be verified that $({\listenfrac{}}', {\xmitfrac{}}')$ satisfies: (i) ${\listenfrac{}}' \listencost{} + {\xmitfrac{}}' \xmitcost{} = \energybound{}$ and ${\listenfrac{}}' = (\numnode-1) {\xmitfrac{}}'$, and (ii) $({\UB})' = \numnode \listenfrac{}' > \UB$.

\noindent{\bf Case 2}: If $\listenfrac{}^{*} \listencost{} + \xmitfrac{}^{*} \xmitcost{} = \energybound{}$ and $\listenfrac{}^{*} = (\numnode-1) \xmitfrac{}^{*} - \delta$ for some $\delta>0$, the optimal solution is given by $\xmitfrac{}^{*} = \frac{\energybound{} + \delta\listencost{}}{\xmitcost{} + (\numnode-1) \listencost{}}, \listenfrac{}^{*} = (\numnode-1) \xmitfrac{}^{*} - \delta$. Let a new solution be
\begin{align*}
\listenfrac{} '' = \listenfrac{}^{*} + \frac{\delta \xmitcost{}}{\xmitcost{} + (\numnode-1) \listencost{}}, ~\xmitfrac{}'' = \xmitfrac{}^{*} - \frac{\delta \listencost{}}{\xmitcost{} + (\numnode-1) \listencost{}},
\end{align*}
and it can be verified that $({\listenfrac{}}'', {\xmitfrac{}}'')$ satisfies: (i) ${\listenfrac{}}'' \listencost{} + {\xmitfrac{}}'' \xmitcost{} = \energybound{}$ and ${\listenfrac{}}'' = (\numnode-1) {\xmitfrac{}}''$, (ii) ${\xmitfrac{}}'' > 0$ still holds, and (iii) $({\UB})'' = \numnode \listenfrac{}'' > \UB$.

Next, given that equalities strictly hold in constraints~\eqref{eqn:energy-constraint} and~\eqref{eqn:listen-when-xmit}, the optimal solution of $(\listenfrac{}^{*}, \xmitfrac{}^{*})$ can be obtained by solving $\listenfrac{}^{*} \listencost{} + \xmitfrac{}^{*} \xmitcost{} = \energybound{}$ and $\listenfrac{}^{*} = (\numnode-1) \xmitfrac{}^{*}$, and the solution is described in Section~\ref{ssec:upper-groupput}. Similarly, {\namelpaput} can be solved to obtain the results described in Section~\ref{ssec:upper-anyput}. \hfill $\blacksquare$

\section{Proof of Lemma~\ref{lem:dbe}}
\label{append:proof-dbe}

For the capture version {\namecap}, we prove that the transition rates~\eqref{eqn:trans-rate-sl}, \eqref{eqn:trans-rate-ls}, \eqref{eqn:trans-rate-lx-cap}, and~\eqref{eqn:trans-rate-xl-cap} will drive the network Markov chain to a steady state with distribution~\eqref{eqn:steady-state-prob}, by checking that the {\em detailed balanced equations} hold.
We assume $\oursigma = 1$ and drop the constant term $1/\oursigma$ for brevity. For network state $\statevec$, define $\setnodes_s(\statevec)$, $\setnodes_l(\statevec)$, and $\setnodes_x(\statevec)$ as the sets of nodes in sleep, listen, and transmit states, respectively, and their cardinalities as $\numnode_s(\statevec)$, $\numnode_l(\statevec)$, and $\numnode_x(\statevec)$, respectively. Note that $\numnode_s(\statevec) + \numnode_l(\statevec) + \numnode_x(\statevec) = \numnode$ and $\numnode_x(\statevec) \in \{0,1\}$. We also use $\statevec = (\setnodes_s(\statevec), \setnodes_l(\statevec), \setnodes_x(\statevec))$ to denote network state $\statevec$ and $r(\statevec, \statevec')$ to denote the transition rate from state $\statevec$ to $\statevec'$. We consider the following cases.

\noindent{\bf Case 1}: If node $i$ is in sleep state ($\state_i = s$), the only transition that can happen is to transition into listen state when the channel is clear, i.e., $\statevec = (\setnodes_s(\statevec), \setnodes_l(\statevec), \varnothing ) \to \statevec' = (\setnodes_s(\statevec) \setminus \{i\}, \setnodes_l(\statevec) \cup \{i\}, \varnothing )$.
In this case, $r(\statevec, \statevec') = \exp( -\mplier_i \listencost{i})$ and $r(\statevec', \statevec) = 1$.

\noindent{\bf Case 2}: If node $i$ is in listen state ($\state_{i} = l$), and transitions to sleep state, i.e., $\statevec = (\setnodes_s(\statevec), \setnodes_l(\statevec), \varnothing ) \to \statevec' = (\setnodes_s(\statevec) \cup \{i\}, \setnodes_l(\statevec) \setminus \{i\}, \varnothing )$.
In this case, $r(\statevec, \statevec') = 1$ and $r(\statevec', \statevec) = \exp( -\mplier_i \listencost{i})$.

\noindent{\bf Case 3}: If node $i$ is in listen state ($\state_{i} = l$) and transitions to transmit state, i.e., $\statevec = (\setnodes_s(\statevec), \setnodes_l(\statevec), \varnothing ) \to \statevec' = (\setnodes_s(\statevec), \setnodes_l(\statevec) \setminus \{i\}, \{i\} )$.
In this case, $r(\statevec, \statevec') = \exp( \mplier_i (\listencost{i} - \xmitcost{i}) )$ and $r(\statevec', \statevec) = \exp( -\numnode_l(\statevec'))$.

\noindent{\bf Case 4}: If node $i$ is in transmit state ($\state_{i} = x$), the only transition that can happen is to transition to listen state when its transmission is finished, i.e., $\statevec = (\setnodes_s(\statevec), \setnodes_l(\statevec), \{i\} ) \to \statevec' = (\setnodes_s(\statevec) \{i\}, \setnodes_l(\statevec) \cup \{i\}, \varnothing )$.
In this case, $r(\statevec, \statevec') = \exp( -\numnode_l(\statevec))$ and $r(\statevec', \statevec) = \exp( \mplier_i(\listencost{i} - \xmitcost{i}) )$.

For each case, $\steadystate_{\statevec} \cdot r(\statevec, \statevec') = \steadystate_{\statevec'} \cdot r(\statevec', \statevec)$ holds and similar detailed balance equations hold for the non-capture version {\namenoncap}. Therefore, we complete the proof. \hfill $\blacksquare$

\section{Proof of Part (iii)}
\label{append:proof-convergence}


We use the same notation described in Appendix~\ref{append:proof-dbe}. In addition, denote $\numstate = \onenorm{\statespace}$ as the number of network states. Let $t_{0} = 0$ and recall that the length of the $k$-th interval is $\epoch_{k} = t_{k} - t_{k-1}$. At time $t$, the probability of the system being in state $\statevec$ is denoted by $\networkprob_{\statevec}(t)$. Notice that throughout the $k$-th interval, the vector of Lagrange multipliers $\mpliervec(k-1)$ remains unchanged and is updated at time $t_{k}$ according to
\begin{align}
\label{eqn:proof-update-lm-emp}
\mplier_{i}(k) = [ \mplier_{i}(k-1) - \stepsize_{k} ( \powerharv_{i} - \powerconsemp_{i}(k) ) ]^+, ~\forall i \in \setnodes
\end{align}
in which $\powerconsemp_i(k)$ is the \textit{empirical} energy consumption rate of node $i$ in the $k$-th interval.\footnote{Although we assume a constant power budget of $\powerharv_i$ at each node $i$, the proof can be easily extended to scenarios where $\powerharv_i$ varies with time.}
The relationship between the empirical energy consumption rate and the energy storage level $\bat_{i}(t)$ of node $i$ is given by $\powerharv_{i} - \powerconsemp_{i}(k) = [ \bat_{i}(t_{k}) - \bat_{i}(t_{k-1}) ] / \epoch_{k}$.

In addition, let $\statevec^0(k)$ denote the state of the network Markov chain at the beginning of the $k$-th interval (i.e., at time $t_{k-1}$). Define the random vector $\mathbf{U}(k-1) := [ \powerconsempvec(k-1), \mpliervec(k-1), \statevec^0(k-1) ]$. For $k \geq 1$, let $\markovhist_{k-1}$ be the $\sigma$-field generated by $\mathbf{U}(0), \mathbf{U}(1), \cdots, \mathbf{U}(k-1)$.

In the $k$-th interval, define the gradient vector as $\gradientvec(k) = [\gradient_i(k)]$, in which $\gradient_i(k) = \powerharv_i(k) - \powercons_i(k)$. Given vector $\mpliervec(k-1)$, $\gradientvec(k)$ is a gradient of $\mathcal{L}(\mpliervec)$ (recall that the dual problem of {\namecp} is $\min\nolimits_{\mpliervec \succeq \bm{0}} \mathcal{L}(\mpliervec)$). 
However, {\name} follows~\eqref{eqn:proof-update-lm-emp} and only has an empirical estimation $\gradientempvec(k) = [\gradientemp_i(k)]$ of $\gradientvec(k)$, in which $\gradientemp_i(k) = \powerharv_i - \powerconsemp_i(k)$. The ``error'' term is given by $\gradientvec^{\rm err}(k) = [\gradient_i^{\rm err}(k)]$, in which
\begin{align}
\label{eqn:proof-gradient-err}
\gradient_i^{\rm err}(k) & = \expectation{\gradientemp_i(k) | \markovhist_{k-1}} - \gradient_i(k) = \powercons_i(k) - \expectation{ \powerconsemp_i(k) | \markovhist_{k-1} }.
\end{align}
The ``noise'' term is given by $\gradientvec^{\rm noise}(k) = [\gradient_i^{\rm noise}(k)]$, in which
\begin{align}
\label{eqn:proof-gradient-noise}
\textstyle \gradient_i^{\rm noise}(k) = \expectation{ \powerconsemp_i(k) | \markovhist_{k-1} } -  \powerconsemp_i(k).
\end{align}
Therefore we can write the empirical gradient as
\begin{align*}
\gradientemp_{i}(k) = \gradient_{i}(k) + \gradient_{i}^{\rm err}(k) + \gradient_{i}^{\rm noise}(k), ~\forall i \in \setnodes.
\end{align*}
Notice that since $\powerharv_i(k)$ and $\powerconsemp_i(k)$ are both bounded, the noise term is also bounded by some constant, i.e., $\onenorm{\gradient_i^{\rm noise}(k)} \leq \gradientmax^{\rm noise}, ~\forall i \in \setnode$.
Below, we show that with {\name}, the error term $\gradientvec^{\rm err}(k)$~\eqref{eqn:proof-gradient-err} decreases ``fast enough'' with time.
We focus on the $(k+1)$-th interval and denote the Continuous-Time Markov chain (CTMC) in the $(k+1)$-th interval by $X(t),~\forall t \in [t_k, t_{k+1})$, whose transition rate matrix is denoted as ${\bf{Q}} = [Q(\statevec, \statevec')]$. For brevity, we assume $\oursigma=1$ and drop the term $1/\oursigma$. Recall from~\eqref{eqn:steady-state-prob}, $Z^{\mpliervec(k)}$ is a normalizing constant which can be easily bounded by
\begin{align*}
Z^{\mpliervec(k)}
\leq \numstate \cdot \exp{(\numnode)} = (2+\numnode) \cdot 2^{\numnode-1} \cdot \exp(\numnode),
\end{align*}
in which we use the fact that $\tput_{\statevec} \leq \numnode$. On the other hand, denote $\costmax = \max\nolimits_{i \in \setnodes} \left\{ \max \left\{ \listencost{i}, \xmitcost{i} \right\} \right\}$, we also have
\begin{align*}
\steadystate_{\statevec}^{\mpliervec(k)} \cdot Z^{\mpliervec(k)}
\geq \exp [ - \numnode \cdot \costmax \cdot \max\limits_i \{ \mplier_i(k) \} ].
\end{align*}
From~\eqref{eqn:proof-update-lm-emp}, $\max\nolimits_i \{ \mplier_i(k) \}$ can be further bounded by $\max\nolimits_{i \in \setnodes} \{\mplier_{i}(k)\} \leq \costmax \cdot \sum\nolimits_{m=1}^{k} \stepsize_{m}$. Therefore, denoting $\xi_{k} := \costmax^{2} \cdot \sum\nolimits_{m=1}^{k} \stepsize_{m}$, we have the minimum probability in the stationary distribution lower bounded by
\begin{align}
\label{eqn:proof-steadystate-min}
\steadystate_{\statevec}^{\mpliervec(k)} \geq \exp ( - \xi_{k} \numnode )  \left( \numstate \cdot \exp(\numnode) \right)^{-1} := \steadystate_{\rm min}^{\mpliervec(k)}.
\end{align}

Next, we state the following useful proposition.

\begin{proposition}
For the Markov chain $X(t)$ in the $(k+1)$-th interval, if $Q(\statevec, \statevec') > 0$, then there exists $Q_{k+1}^{\rm max} \geq Q_{k+1}^{\rm min} > 0$ such that $Q_{k+1}^{\rm min} \leq Q(\statevec, \statevec') \leq Q_{k+1}^{\rm max}$.
\end{proposition}
\begin{IEEEproof}
Given network state $\statevec$, there is $\numnode_s(\statevec) + 2\numnode_l(\statevec) + \numnode_x(\statevec) \leq 2\numnode$ states $\statevec'$ other than $\statevec$ that $X(t)$ can transition to. For any state $\statevec' \ne \statevec$, we have for each node $i$,
\begin{align}
\label{eqn:trans-rate-matrix}
& Q(\statevec, \statevec') = \left\{
\begin{aligned}
& \exp [ - \mplier_i(k) \listencost{i} ], && \state_{i}: s \to l, \\
& \exp [ \mplier_i(k) (\listencost{i} - \xmitcost{i}) ], && \state_{i}: l \to x, \\
& 1, && \state_{i}: l \to s, \\
& \exp( - \tput_{\statevec}), && \state_{i}: x \to l.
\end{aligned}
\right.
\end{align}
It is easy to see that $Q_{k+1}^{\rm min} = \min\{ \exp(-\numnode), \exp(-\xi_k) \}$ suffices. In particular, given the network size $\numnode$, for sufficiently large $k$, $\xi_k > \numnode$ holds, and therefore we use $Q_{k+1}^{\rm min} := \exp(-\xi_{k})$.

On the other hand, if for $\listencost{i} \leq \xmitcost{i}, \forall i \in \setnodes$, each one of the $2\numnode$ transition rates in~\eqref{eqn:trans-rate-matrix} is less than or equal to $1$, and $Q_{k+1}^{\rm max} = 2\numnode$ is sufficient. Second, if there exists some node $i$ such that $\listencost{i} > \xmitcost{i}$, beacause of the costs are always upper bounded by $\costmax$, the transition rate is also upper bounded by $\exp(\costmax \mpliermax) \leq \exp{(\costmax^{2} \sum\nolimits_{m=1}^{k} \stepsize_{m})} = \exp{(\xi_{k})}$. Therefore $Q_{k+1}^{\rm max} = 2\numnode \exp{(\xi_{k})}$ is sufficient.
\end{IEEEproof}

Following the standard method, we perform uniformization on $X(t)$ whose transition rate matrix is denoted as $\bf{Q}$. If each element of $\bf{Q}$ has an absolulte value less that a constant $Q_{k+1}^{\rm max}$, then we can write $X(t) = Y(M(t))$, in which $Y(n)$ is a discrete time Markov chain with probability transition matrix ${\bf{P}} = {\bf{I}} + {\bf{Q}} / Q_{k+1}^{\rm max}$ and $\bf{I}$ is the identity matrix. $M(t)$ is an independent Poisson process with rate $Q_{k+1}^{\rm max}$.


Then, we estimate how far the empirical power consumption $\expectation{\powerconsemp_i(k+1) | \markovhist_k}$ is away from the desired value $\powercons_{i}(k+1)$ (under fixed $\mpliervec(k)$). Denote the probability of $X(t)$ in state $\statevec$ at time $t \in [t_{k}, t_{k+1})$ by $\networkprob_{\statevec}(t)$ and $\networkprobvec(t) = [\networkprob_{\statevec}(t)]$. Given the initial state at time $t_k$ is $\statevec^0(k+1)$, We have
\begin{align*}
& \textstyle \expectation{ \powerconsemp_i(k+1) | \markovhist_k }\\
& \textstyle = \expectation{ \int_{t_k}^{t_{k+1}} \sum_{\statevec \in \statespace} \left( \mathbf{1}_{
\{\state_i = l\} } \listencost{i} +  \mathbf{1}_{ \{ \state_i = x \} } \xmitcost{i} \right)\, \textrm{d}t / \epoch_{k+1} } \\
& \textstyle = \int_{t_k}^{t_{k+1}} \sum_{\statevec \in \statespace} ( \prob \left\{ \state_i(t) = l \right\} \listencost{i} + \prob \left\{ \state_i(t) = x \right\} \xmitcost{i} )\, \textrm{d}t / \epoch_{k+1} \\
& \textstyle = \frac{\listencost{i}}{\epoch_{k+1}} \sum_{\statevec \in \statespace_i^l} \int_{t_k}^{t_{k+1}} \networkprob_{\statevec}(t)\, \textrm{d}t + \frac{\xmitcost{i}}{\epoch_{k+1}} \sum_{\statevec \in \statespace_i^x} \int_{t_{k}}^{t_{k+1}} \networkprob_{\statevec}(t)\, \textrm{d}t \\
& \textstyle = \listencost{i} \cdot \listenfracemp{i}(k+1) + \xmitcost{i} \cdot \xmitfracemp{i}(k+1),
\end{align*}
in which
\begin{align*}
& \textstyle \emplistenfrac{i}(k+1) = \sum\nolimits_{\statevec \in \statespace_i^l} \int_{t_{k}}^{t_{k+1}} \networkprob_{\statevec}(t) \, \textrm{d}t / \epoch_{k+1}, \\
& \textstyle \xmitfracemp{i}(k+1) = \sum\nolimits_{\statevec \in \statespace_i^x} \int_{t_{k}}^{t_{k+1}} \networkprob_{\statevec}(t) \, \textrm{d}t / \epoch_{k+1},
\end{align*}
are the empirical average probabilities that node $i$ spends in listen and transmit states in the $(k+1)$-th interval.

Let $\slev$ be the Second Largest Eigenvalue modulus of the transition probability matrix $\bm{P} = \bm{I} + \bm{Q} / Q_{k+1}^{\rm max}$.
By Theorem~\cite{}, this total variation distance can be bounded by
\begin{align}
\label{eqn:proof-norvtv-bound}  
\textstyle \normtv{\networkprobvec(t) - \steadystatevec^{\mpliervec(k)}}
& \textstyle = \frac{1}{2} \sum\nolimits_{\statevec \in \statespace} \onenorm{ \networkprob_{\statevec}(t) - \steadystate_{\statevec}^{\mpliervec(k)} } \nonumber \\
& \textstyle \leq \frac{1}{2} \sqrt{ \frac{1 - \steadystate_{\statevec^{0}}^{\mpliervec(k)}}{\steadystate_{\statevec^{0}}^{\mpliervec(k)}} } \exp ( - Q_{k+1}^{\rm max} (1-\slev) t) \nonumber \\
& \textstyle \leq \frac{1}{2} \frac{1}{\sqrt{\steadystate_{\rm min}^{\mpliervec(k)}}} \exp ( - Q_{k+1}^{\rm max} (1-\slev) t).
\end{align}
Also, the Second Largest Eigenvalue, $\slev$, can be bounded by Cheeger's inequality~\cite{diaconis1991geometric}, i.e.,
\begin{align*}
\slev \leq 1 - \phi^{2} / 2 \Leftrightarrow 1 / (1-\theta_{2}) \leq 2 / \phi^{2},
\end{align*}
where $\phi$ is the ``conductance'' of $\bm{P}$, which satisfies the following inequality
\begin{align*}
\textstyle \phi & \geq \min_{\statevec \ne \statevec', P(\statevec, \statevec') > 0} \{  \steadystate_{\rm min}^{\mpliervec(k)} P(\statevec, \statevec') \} \geq \steadystate_{\rm min}^{\mpliervec(k)} \cdot Q_{k+1}^{\rm min} / Q_{k+1}^{\rm max}.
\end{align*}
Therefore we obtain
\begin{align}
\label{eqn:proof-slev-bound}
1 / (1-\slev) \leq 2 / \phi^{2} \leq 2 (Q_{k+1}^{\rm max})^{2}/
( \steadystate_{\rm min}^{\mpliervec(k)} \cdot Q_{k+1}^{\rm min})^{2}.
\end{align}

Putting everything into the estimation, we have
\begin{align*}
& \onenorm{ \expectation{ \powerconsemp_i(k+1) | \markovhist_k } - \powercons_i(k+1)} \\
& = \onenorm{ \listencost{i} (\listenfracemp{i}(k+1) - \listenfrac{i}(k+1)) + \xmitcost{i} (\xmitfracemp{i}(k+1) - \xmitfrac{i}(k+1)) } \\
& \leq \listencost{i} \onenorm{ \listenfracemp{i}(k+1) - \listenfrac{i}(k+1) } + \xmitcost{i} \onenorm{ \xmitfracemp{i}(k+1) - \xmitfrac{i}(k+1) }.
\end{align*}
Furthermore, we have
\begin{align*}
& \textstyle \onenorm{ \listenfracemp{i}(k+1) - \listenfrac{i}(k+1) } \\
& \textstyle  = \onenorm{ \sum\nolimits_{\statevec \in \statespace_i^l} \int_{t_{k}}^{t_{k+1}} ( \networkprob_{\statevec}(t) - \steadystate_{\statevec}^{\mpliervec(k)} ) \, \textrm{d}t / \epoch_{k+1} } \\
& \textstyle \leq \int_{t_{k}}^{t_{k+1}} \sum\nolimits_{\statevec \in \statespace_i^l} \onenorm{ \networkprob_{\statevec}(t) - \steadystate_{\statevec}^{\mpliervec(k)} } \, \textrm{d}t / \epoch_{k+1}.
\end{align*}
Similarly, $ \onenorm{ \xmitfracemp{i}(k+1) - \xmitfrac{i}(k+1) } \leq \int_{t_{k}}^{t_{k+1}} \sum\nolimits_{\statevec \in \statespace_i^x} \onenorm{ \networkprob_{\statevec}(t) - \steadystate_{\statevec}^{\mpliervec(k)} } \, \textrm{d}t / \epoch_{k+1}$.
Adding both terms together yields
\begin{align*}
& \textstyle \listencost{i} \onenorm{ \listenfracemp{i}(k+1) - \listenfrac{i}(k+1) } + \xmitcost{i} \onenorm{ \xmitfracemp{i}(k+1) - \xmitfrac{i}(k+1) } \\
& \textstyle \leq \costmax \int_{t_k}^{t_{k+1}} \sum\nolimits_{\statevec \in \statespace} \onenorm{ \networkprob_{\statevec}(t) - \steadystate_{\statevec}^{\mpliervec(k)} } \, \textrm{d}t / \epoch_{k+1} \\
& \textstyle \leq \frac{\costmax}{\sqrt{\steadystate_{\rm min}^{\mpliervec(k)}}} \int_{t_k}^{t_{k+1}} \exp{(-Q_{k+1}^{\rm max}(1-\slev)t)} \, \textrm{d}t / \epoch_{k+1} ~\textrm{(use~\eqref{eqn:proof-norvtv-bound})} \\
& \textstyle \leq \frac{\costmax}{\sqrt{\steadystate_{\rm min}^{\mpliervec(k)}}} \int_{0}^{t_{k+1}} \exp{(-Q_{k+1}^{\rm max}(1-\slev)t)} \, \textrm{d}t / \epoch_{k+1} \\
& \textstyle \leq \frac{\costmax}{\sqrt{\steadystate_{\rm min}^{\mpliervec(k)}}} \cdot \left[ 2 Q_{k+1}^{\rm max} (1-\slev) \right]^{-1} / \epoch_{k+1} \\
& \textstyle \leq \frac{\costmax Q_{k+1}^{\rm max}}{ \epoch_{k+1} (Q_{k+1}^{\rm min})^2 (\steadystate_{\rm min}^{\mpliervec(k)})^{5/2} }
:= f(k) / \epoch_{k+1}, ~\textrm{(use~\eqref{eqn:proof-slev-bound})}
\end{align*}
in which $f(k) := \frac{\costmax Q_{k+1}^{\rm max}}{ (Q_{k+1}^{\rm min})^2 (\steadystate_{\rm min}^{\mpliervec(k)})^{5/2} }$.
Therefore we have,
\begin{align*}
\onenorm{ \expectation{ \powerconsemp_i(k+1) | \markovhist_k } - \powercons_i(\mpliervec(k)) } \leq f(k) / \epoch_{k+1},
\end{align*}
and the error term, defined in~\eqref{eqn:proof-gradient-err}, satisfies $\onenorm{ \gradient_i^{\rm err}(k+1)} \leq f(k) / \epoch_{k+1}$. Therefore, with defined $f(k)$ and proper choice of $\epoch_{k+1}$ (e.g., $\epoch_{k+1} = k+1$), the error term is diminishing.

Based on this result and following similar steps in the proof of Theorem 1 in~\cite{jiang2009convergence}, it can be shown that $\mpliervec$ converges to $\mpliervec^{*}$ with probability $1$. Therefore, we complete the proof. \hfill $\blacksquare$

\section{Burstiness Analysis of {\name}}
\label{append:analysis-burstiness}
To derive the average burst length (denoted by $B$) of {\namecap}, we use $\steadystate_{\statevec}^{*}$ to denote the optimal solution to {\namecp} and define $\statespace' = \left\{ \statevec \in \statespace: \indionexmit{\statevec} = 1, \numlisten_{\statevec} \geq 1 \right\}$, i.e., the set of states with successfully received bursts. Recall that for a given value of $\oursigma$, the optimal value of {\namecp} is exactly $\tputsigma$.
According to~\eqref{eqn:trans-rate-xl-cap}, for a given state $\statevec \in \statespace'$, the average burst length of {\namecap} in groupput mode is $\exp{(\numlisten_{\statevec}/\oursigma)}$.
Therefore, during a (long enough) time duration of $T$, the average number of bursts received by all the nodes can be computed by $\sum_{\statevec \in \statespace'} \frac{ T \cdot \steadystate_{\statevec}^{*} }{ \exp{(\numlisten_{\statevec}/\oursigma)} }$, and the average burst length of {\name} in groupput mode, $B_{g}$, is given by
\begin{align}
B_{g} & = \textrm{(Avg. Total Burst Length)} / \textrm{(Avg. Number of Bursts)} \nonumber \\
& = \frac{ T \cdot \sum_{\statevec \in \statespace'} \steadystate_{\statevec}^{*} }{ \sum_{\statevec \in \statespace'} \frac{ T \cdot \steadystate_{\statevec}^{*} }{ \exp{(\numlisten_{\statevec}/\oursigma)} } }
= \frac{ \sum_{\statevec \in \statespace'} \steadystate_{\statevec}^{*} }{ \sum_{\statevec \in \statespace'} \frac{ \steadystate_{\statevec}^{*} }{ \exp{(\numlisten_{\statevec}/\oursigma)} } }.
\label{eqn:analysis-burstiness-gput}
\end{align}
Similarly, the average burst length of {\namecap} in anyput mode is computed by replacing $\numlisten_{\statevec}$ with $\indisomelisten_{\statevec}$ in~\eqref{eqn:analysis-burstiness-gput}. Since $\indisomelisten_{\statevec} = 1$ always holds for $\statevec \in \statespace'$, $B_{a}$ simplifies to
\begin{align}
B_{a} & = \frac{ \sum_{\statevec \in \statespace'} \steadystate_{\statevec}^{*} }{ \sum_{\statevec \in \statespace'} \left[ \steadystate_{\statevec}^{*} \cdot \exp{(-1/\oursigma)} \right] } = \exp{(1/\oursigma)}.
\label{eqn:analysis-burstiness-aput}
\end{align}
This shows that the anyput average burst length is independent of the number of nodes, $\numnode$, and only depends on $\oursigma$. \hfill $\blacksquare$



\end{document}